\pdfoutput=1
\documentclass[12pt,a4paper]{article}
\usepackage{ifthen} % for conditional statements
\usepackage{rotating}
\usepackage{dashrule}
\usepackage{array} 
\usepackage{dcolumn}
\usepackage{comment}
\usepackage{xcolor}

\usepackage{graphpap} 
\usepackage{rotating} 
\usepackage{tikz}
\usepackage{xcolor}
\usepackage{longtable} % only for template; not usually to be used in PAPERs
\usepackage[normalem]{ulem}

\usepackage{lscape}
\usepackage{multirow} 
\usepackage{mathrsfs}
\usepackage{mathtools} %% dcases 
\usetikzlibrary{patterns}
\usepackage{nicefrac} %% 1/2, ...
\usepackage{transparent}
\usepackage{afterpage}

\newboolean{pdflatex}
\setboolean{pdflatex}{true} % False for eps figures 

\newboolean{articletitles}
\setboolean{articletitles}{true} % False removes titles in references

\newboolean{uprightparticles}
\setboolean{uprightparticles}{True} %True for upright particle symbols

\def\paperauthors{LHCb collaboration} % Leave as is for PAPER and CONF
\def\paperasciititle{Study of the the psi2(3823) and chic1(3872) states in 
B+ -> (J/psi pi+ pi-)K+ decays} % Set ASCII title here
\def\papertitle{Study of the~\psitwod and~\chiconex~states 
in~\mbox{$\decay{\Bu}{\left(\jpsi\pip\pim\right)\Kp}$} decays} % Latex formatted title
\def\paperkeywords{{High Energy Physics}, {LHCb}} % Comma separated list
\def\papercopyright{\the\year\ CERN for the benefit of the LHCb collaboration} % new since 9/Apr/2018
\def\paperlicence{CC BY 4.0 licence}
\def\paperlicenceurl{https://creativecommons.org/licenses/by/4.0/}

\makeatletter
\g@addto@macro\bfseries{\boldmath}
\makeatother

\usepackage[top=1in, bottom=1.25in, left=1in, right=1in]{geometry}

\usepackage{microtype}
\usepackage{lineno}  % for line numbering during review
\usepackage{xspace} % To avoid problems with missing or double spaces after
\usepackage{caption} %these three command get the figure and table captions automatically small

\usepackage{graphicx}  % to include figures (can also use other packages)
\usepackage{color}
\usepackage{colortbl}
\graphicspath{{./figs/}} % Make Latex search fig subdir for figures
\DeclareGraphicsExtensions{.pdf,.PDF,png,.PNG}

\usepackage{amsmath} % Adds a large collection of math symbols
\usepackage{amssymb}
\usepackage{amsfonts}
\usepackage{upgreek} % Adds in support for greek letters in roman typeset

\newcommand*\patchAmsMathEnvironmentForLineno[1]{%
\expandafter\let\csname old#1\expandafter\endcsname\csname #1\endcsname
\expandafter\let\csname oldend#1\expandafter\endcsname\csname
end#1\endcsname
 \renewenvironment{#1}%
   {\linenomath\csname old#1\endcsname}%
   {\csname oldend#1\endcsname\endlinenomath}%
}
\newcommand*\patchBothAmsMathEnvironmentsForLineno[1]{%
  \patchAmsMathEnvironmentForLineno{#1}%
  \patchAmsMathEnvironmentForLineno{#1*}%
}
\AtBeginDocument{%
\patchBothAmsMathEnvironmentsForLineno{equation}%
\patchBothAmsMathEnvironmentsForLineno{align}%
\patchBothAmsMathEnvironmentsForLineno{flalign}%
\patchBothAmsMathEnvironmentsForLineno{alignat}%
\patchBothAmsMathEnvironmentsForLineno{gather}%
\patchBothAmsMathEnvironmentsForLineno{multline}%
\patchBothAmsMathEnvironmentsForLineno{eqnarray}%
}

\usepackage{hyperxmp}

\usepackage[pdftex,
            pdfauthor={\paperauthors},
            pdftitle={\paperasciititle},
            pdfkeywords={\paperkeywords},
            pdfcopyright={Copyright (C) \papercopyright},
            pdflicenseurl={\paperlicenceurl}]{hyperref}

\usepackage[colorinlistoftodos,textsize=scriptsize]{todonotes}

\usepackage[all]{hypcap} % Internal hyperlinks to floats.

\setboolean{uprightparticles}{true} %Set true for upright particle symbols
\usepackage{xspace} 
\usepackage{upgreek}

\def\lhcb   {\mbox{LHCb}\xspace}

\def\belle  {\mbox{Belle}\xspace}

\def\besiii {\mbox{BES\,III}\xspace}

\def\MagUp {\mbox{\em Mag\kern -0.05em Up}\xspace}

\ifthenelse{\boolean{uprightparticles}}%
{
 
 \def\Pgamma      {\ensuremath{\upgamma}\xspace}                 
 \def\Pdelta      {\ensuremath{\updelta}\xspace}

 \def\Pmu         {\ensuremath{\upmu}\xspace}

 \def\Ppi         {\ensuremath{\uppi}\xspace}                 
                  
 \def\Prho        {\ensuremath{\uprho}\xspace}

 \def\Pphi        {\ensuremath{\upphi}\xspace}                 
                  
 \def\Pchi        {\ensuremath{\upchi}\xspace}                 
 \def\Ppsi        {\ensuremath{\uppsi}\xspace}

 \def\PDelta      {\ensuremath{\Delta}\xspace}                 
 \def\PXi         {\ensuremath{\Xi}\xspace}                 
 \def\PLambda     {\ensuremath{\Lambda}\xspace}                 
 \def\PSigma      {\ensuremath{\Sigma}\xspace}                 
 \def\POmega      {\ensuremath{\Omega}\xspace}                 
 \def\PUpsilon    {\ensuremath{\Upsilon}\xspace}

 \def\PB      {\ensuremath{\mathrm{B}}\xspace}                 
                  
 \def\PD      {\ensuremath{\mathrm{D}}\xspace}

 \def\PJ      {\ensuremath{\mathrm{J}}\xspace}                 
 \def\PK      {\ensuremath{\mathrm{K}}\xspace}

 \def\PX      {\ensuremath{\mathrm{X}}\xspace}

 \def\Pb      {\ensuremath{\mathrm{b}}\xspace}                 
 \def\Pc      {\ensuremath{\mathrm{c}}\xspace}

 \def\Pi      {\ensuremath{\mathrm{i}}\xspace}

 \def\Pp      {\ensuremath{\mathrm{p}}\xspace}

 \def\Ps      {\ensuremath{\mathrm{s}}\xspace}

 \def\thebaroffset{0.0em}
}
{
 
 \def\Pgamma      {\ensuremath{\gamma}\xspace}                 
 \def\Pdelta      {\ensuremath{\delta}\xspace}

 \def\Pmu         {\ensuremath{\mu}\xspace}

 \def\Ppi         {\ensuremath{\pi}\xspace}                 
                  
 \def\Prho        {\ensuremath{\rho}\xspace}

 \def\Pphi        {\ensuremath{\phi}\xspace}                 
                  
 \def\Pchi        {\ensuremath{\chi}\xspace}                 
 \def\Ppsi        {\ensuremath{\psi}\xspace}                 
                  
 \mathchardef\PDelta="7101
 \mathchardef\PXi="7104
 \mathchardef\PLambda="7103
 \mathchardef\PSigma="7106
 \mathchardef\POmega="710A
 \mathchardef\PUpsilon="7107
                  
 \def\PB      {\ensuremath{B}\xspace}                 
                  
 \def\PD      {\ensuremath{D}\xspace}

 \def\PJ      {\ensuremath{J}\xspace}                 
 \def\PK      {\ensuremath{K}\xspace}

 \def\PX      {\ensuremath{X}\xspace}

 \def\Pb      {\ensuremath{b}\xspace}                 
 \def\Pc      {\ensuremath{c}\xspace}

 \def\Pi      {\ensuremath{i}\xspace}

 \def\Pp      {\ensuremath{p}\xspace}

 \def\Ps      {\ensuremath{s}\xspace}

 \def\thebaroffset{0.18em}
}
\newcommand{\offsetoverline}[2][\thebaroffset]{\kern #1\overline{\kern -#1 #2}}%

\makeatletter
\ifcase \@ptsize \relax% 10pt
  \newcommand{\miniscule}{\@setfontsize\miniscule{4}{5}}% \tiny: 5/6
\or% 11pt
  \newcommand{\miniscule}{\@setfontsize\miniscule{5}{6}}% \tiny: 6/7
\or% 12pt
  \newcommand{\miniscule}{\@setfontsize\miniscule{5}{6}}% \tiny: 6/7
\fi
\makeatother

\DeclareRobustCommand{\optbar}[1]{\shortstack{{\miniscule (\rule[.5ex]{1.25em}{.18mm})}
  \\ [-.7ex] $#1$}}

\def\mumu       {{\ensuremath{\Pmu^+\Pmu^-}}\xspace}

\def\g      {{\ensuremath{\Pgamma}}\xspace}

\def\squark    {{\ensuremath{\Ps}}\xspace}

\def\cquark    {{\ensuremath{\Pc}}\xspace}
\def\cquarkbar {{\ensuremath{\overline \cquark}}\xspace}
\def\ccbar     {{\ensuremath{\cquark\cquarkbar}}\xspace}
\def\bquark    {{\ensuremath{\Pb}}\xspace}

\def\pion   {{\ensuremath{\Ppi}}\xspace}

\def\pip    {{\ensuremath{\pion^+}}\xspace}
\def\pim    {{\ensuremath{\pion^-}}\xspace}

\def\kaon    {{\ensuremath{\PK}}\xspace}
\def\KorKbar {\kern \thebaroffset\optbar{\kern -\thebaroffset \PK}{}\xspace}

\def\Kp      {{\ensuremath{\kaon^+}}\xspace}
\def\Km      {{\ensuremath{\kaon^-}}\xspace}

\def\KS      {{\ensuremath{\kaon^0_{\mathrm{S}}}}\xspace}

\def\Dbar    {{\ensuremath{\offsetoverline{\PD}}}\xspace}
\def\D       {{\ensuremath{\PD}}\xspace}

\def\DorDbar {\kern \thebaroffset\optbar{\kern -\thebaroffset \PD}\xspace}
\def\Dz      {{\ensuremath{\D^0}}\xspace}
\def\Dzb     {{\ensuremath{\Dbar{}^0}}\xspace}
\def\Dp      {{\ensuremath{\D^+}}\xspace}
\def\Dm      {{\ensuremath{\D^-}}\xspace}

\def\Dstarz  {{\ensuremath{\D^{*0}}}\xspace}
\def\Dstarzb {{\ensuremath{\Dbar{}^{*0}}}\xspace}

\def\Dstarp  {{\ensuremath{\D^{*+}}}\xspace}

\def\Ds      {{\ensuremath{\D^+_\squark}}\xspace}

\def\B       {{\ensuremath{\PB}}\xspace}

\def\BorBbar {\kern \thebaroffset\optbar{\kern -\thebaroffset \PB}\xspace}

\def\Bd      {{\ensuremath{\B^0}}\xspace}

\def\BdorBdbar {\kern \thebaroffset\optbar{\kern -\thebaroffset \Bd}\xspace}
\def\Bu      {{\ensuremath{\B^+}}\xspace}

\def\Bp      {{\ensuremath{\Bu}}\xspace}

\def\Bs      {{\ensuremath{\B^0_\squark}}\xspace}

\def\BsorBsbar {\kern \thebaroffset\optbar{\kern -\thebaroffset \Bs}\xspace}

\def\jpsi     {{\ensuremath{{\PJ\mskip -3mu/\mskip -2mu\Ppsi\mskip 2mu}}}\xspace}
\def\psitwos  {{\ensuremath{\Ppsi{\rm{(2S)}}}}\xspace}

\def\chicone  {{\ensuremath{\Pchi_{\cquark 1}}}\xspace}
\def\chictwo  {{\ensuremath{\Pchi_{\cquark 2}}}\xspace}

\def\Y#1S{\ensuremath{\PUpsilon{(#1S)}}\xspace}

\def\proton      {{\ensuremath{\Pp}}\xspace}

\def\LorLbar     {\kern \thebaroffset\optbar{\kern -\thebaroffset \PLambda}\xspace}

\def\psitwod{{\ensuremath{\Ppsi_2(3823)}}\xspace}
\def\chiconex{{\ensuremath{\chicone(3872)}}\xspace}

\def\BF         {{\ensuremath{\mathcal{B}}}\xspace}
\def\BR         {\BF}

\newcommand{\decay}[2]{\ensuremath{#1\!\to #2}\xspace} 

\def\to                 {\ensuremath{\rightarrow}\xspace}

\def\eps   {{\ensuremath{\varepsilon}}\xspace}

\def\BpToJpsipipiK {\mbox{\decay{\Bp}{\jpsi\pip\pim\Kp}}}
\def\BpTopsitwosK {\mbox{\decay{\Bp}{\psitwos\Kp}}}

\def\BpTopsiK      {\decay{\Bu}{\Ppsi_2(3823)\Kp}}

\def\BpTopsitwosK  {\decay{\Bu}{\psitwos\Kp}}

\def\BpTochicK     {\decay{\Bu}{\chicone(3872)\Kp}}

\def\ChiconexTojpsipipi{\decay{\chiconex}{\jpsi\pip\pim}}
\def\PsitwodTojpsipipi{\decay{\psitwod}{\jpsi\pip\pim}}
\def\JpsiPiPi{\jpsi\pip\pim}
\def\AT#1     {\ensuremath{A_{\mathrm{T}}^{#1}}\xspace}           % 2

\def\C#1      {\ensuremath{\mathcal{C}_{#1}}\xspace}                       % 9
\def\Cp#1     {\ensuremath{\mathcal{C}_{#1}^{'}}\xspace}                    % 7
\def\Ceff#1   {\ensuremath{\mathcal{C}_{#1}^{\mathrm{(eff)}}}\xspace}        % 9  
\def\Cpeff#1  {\ensuremath{\mathcal{C}_{#1}^{'\mathrm{(eff)}}}\xspace}       % 7
\def\Ope#1    {\ensuremath{\mathcal{O}_{#1}}\xspace}                       % 2
\def\Opep#1   {\ensuremath{\mathcal{O}_{#1}^{'}}\xspace}                    % 7

\newcommand{\nospaceunit}[1]{\ensuremath{\text{#1}}}       
\newcommand{\aunit}[1]{\ensuremath{\text{\,#1}}}       
\newcommand{\tev}{\aunit{Te\kern -0.1em V}\xspace}
\newcommand{\gev}{\aunit{Ge\kern -0.1em V}\xspace}
\newcommand{\mev}{\aunit{Me\kern -0.1em V}\xspace}
\newcommand{\kev}{\aunit{ke\kern -0.1em V}\xspace}
\newcommand{\kevcc}{\ensuremath{\aunit{ke\kern -0.1em V\!/}c^2}\xspace}
\newcommand{\ev}{\aunit{e\kern -0.1em V}\xspace}
\newcommand{\mevc}{\ensuremath{\aunit{Me\kern -0.1em V\!/}c}\xspace}
\newcommand{\gevc}{\ensuremath{\aunit{Ge\kern -0.1em V\!/}c}\xspace}
\newcommand{\mevcc}{\ensuremath{\aunit{Me\kern -0.1em V\!/}c^2}\xspace}
\newcommand{\gevcc}{\ensuremath{\aunit{Ge\kern -0.1em V\!/}c^2}\xspace}
\def\m    {\aunit{m}\xspace}

\def\mum  {\ensuremath{\,\upmu\nospaceunit{m}}\xspace}

\def\fb   {\ensuremath{\aunit{fb}}\xspace}
\def\invfb   {\ensuremath{\fb^{-1}}\xspace}

\newcommand{\chisq}{\ensuremath{\chi^2}\xspace}

\newcommand{\chisqip}{\ensuremath{\chi^2_{\text{IP}}}\xspace}

\def\gsim{{~\raise.15em\hbox{$>$}\kern-.85em
          \lower.35em\hbox{$\sim$}~}\xspace}
\def\lsim{{~\raise.15em\hbox{$<$}\kern-.85em
          \lower.35em\hbox{$\sim$}~}\xspace}

\def\sPlot{\mbox{\em sPlot}\xspace}

\def\pt         {\ensuremath{p_{\mathrm{T}}}\xspace}

\def\evtgen     {\mbox{\textsc{EvtGen}}\xspace}

\def\geant      {\mbox{\textsc{Geant4}}\xspace}

\def\photos     {\mbox{\textsc{Photos}}\xspace}

\def\pythia     {\mbox{\textsc{Pythia}}\xspace}

\def\tell1  {TELL1\xspace}
\def\ukl1   {UKL1\xspace}

\newcommand{\eg}{\mbox{\itshape e.g.}\xspace}

\usepackage{cite} % Allows for ranges in citations
\usepackage{mciteplus}

\usepackage{longtable} % only for template; not usually to be used in PAPERs

\newcolumntype{d}[1]{D{,}{\,\pm\,}{#1} }
\newcolumntype{f}[1]{D{,}{.}{#1} }

\begin{document}

%%%%%%%%%%%%%%%%%%%%%%%%%
%%%%% Title     %%%%%%%%%
%%%%%%%%%%%%%%%%%%%%%%%%%
\renewcommand{\thefootnote}{\fnsymbol{footnote}}
\setcounter{footnote}{1}

% %%%%%%% CHOOSE TITLE PAGE--------
%\onecolumn
% $Id: title-LHCb-PAPER.tex 122889 2018-08-17 17:59:55Z pkoppenb $
% ===============================================================================
% Purpose: LHCb-PAPER journal paper title page template
% Author: 
% Created on: 2010-09-25
% ===============================================================================

%%%%%%%%%%%%%%%%%%%%%%%%%
%%%%%  TITLE PAGE  %%%%%%
%%%%%%%%%%%%%%%%%%%%%%%%%
\begin{titlepage}
\pagenumbering{roman}

% Header ---------------------------------------------------
\vspace*{-1.5cm}
\centerline{\large EUROPEAN ORGANIZATION FOR NUCLEAR RESEARCH (CERN)}
\vspace*{1.5cm}
\noindent
\begin{tabular*}{\linewidth}{lc@{\extracolsep{\fill}}r@{\extracolsep{0pt}}}
\ifthenelse{\boolean{pdflatex}}% Logo format choice
{\vspace*{-1.5cm}\mbox{\!\!\!\includegraphics[width=.14\textwidth]{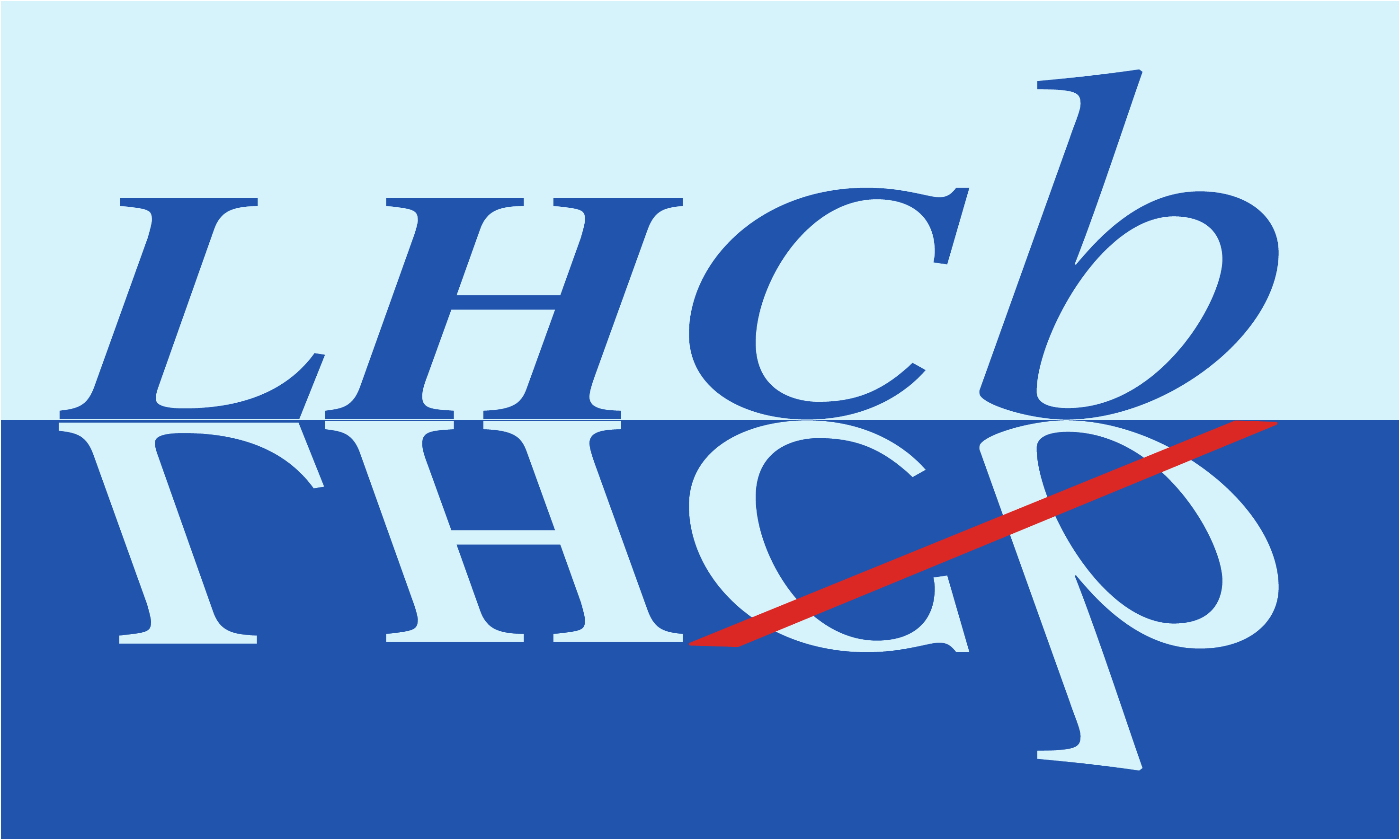}} & &}%
{\vspace*{-1.2cm}\mbox{\!\!\!\includegraphics[width=.12\textwidth]{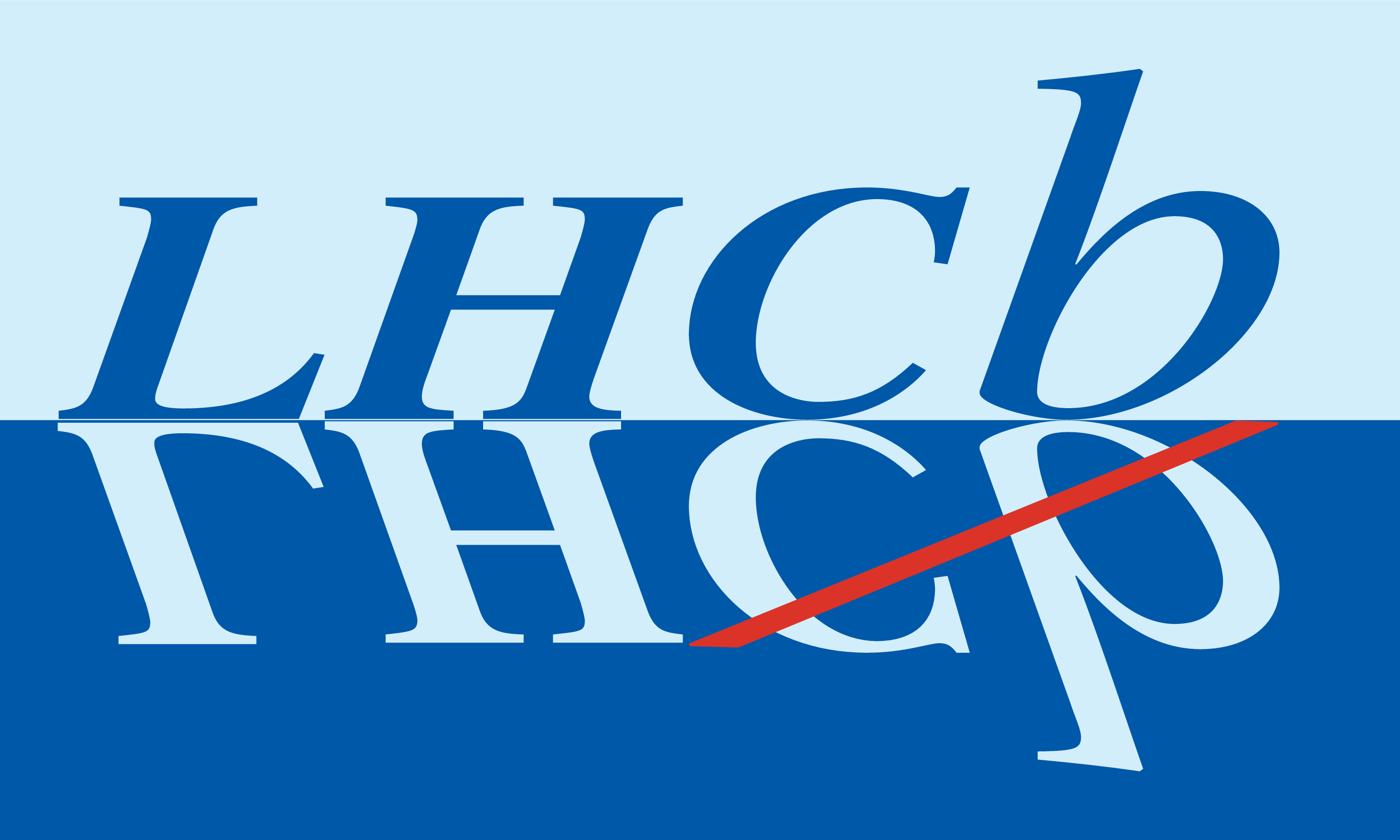}} & &}%
\\
 & & CERN-EP-2020-071\\  % ID 
& & LHCb-PAPER-2020-009 \\  % ID
& & May 27, 2020 \\
%% & & \today \\ % Date - Can also hardwire e.g.: 23 March 2010
%% & & version 0.16 
%% & &
% not in paper \hline
\end{tabular*}

\vspace*{0.2cm}

% Title --------------------------------------------------
{\normalfont\bfseries\boldmath\huge
\begin{center}
% DO NOT EDIT HERE. Instead edit macro in main.tex to keep metadata correct
  \papertitle 
\end{center}
}

\vspace*{0.2cm}

% Authors -------------------------------------------------
\begin{center}
%In the footnote, replace 'paper' by 'Letter' in case of submission to PRL or PLB 
% Edit macro in main.tex to keep metadata correct
\paperauthors\footnote{Authors are listed at the end of this paper.}
\end{center}

\vspace{\fill}

% Abstract -----------------------------------------------
\begin{abstract}
\noindent 
The~decays \mbox{\decay{\Bp}{\jpsi\pip\pim\Kp}}
are studied using a~data set
corresponding to an~integrated luminosity of $9\invfb$ 
collected with the~\lhcb detector in 
proton\nobreakdash-proton collisions between 2011 and 2018.
Precise measurements of the~ratios of 
branching fractions with the~intermediate 
\psitwod, \chiconex and \psitwos~states 
are reported.
The~values are 
\begin{eqnarray*}
\dfrac{\BR_{\decay{\Bu}{\psitwod\Kp}}\times 
       \BR_{\decay{\psitwod}{\jpsi\pip\pim}}}
      {\BR_{\decay{\Bu}{\chiconex\Kp}} \times 
       \BR_{\decay{\chiconex}{\jpsi\pip\pim}}} 
       & = &   \left(3.56 \pm 0.67 \pm 0.11 \right) \times 10^{-2}\,, \\
\dfrac{\BR_{\decay{\Bu}{\psitwod\Kp}} \times 
       \BR_{\decay{\psitwod}{\jpsi\pip\pim}}}
      {\BR_{\decay{\Bu}{\psitwos\Kp}} \times 
       \BR_{\decay{\psitwos}{\jpsi\pip\pim}}} 
       & = &   \left(1.31 \pm 0.25 \pm 0.04 \right) \times 10^{-3}\,, \\
\dfrac{\BR_{\decay{\Bu}{\chiconex\Kp}} \times 
       \BR_{\decay{\chiconex}{\jpsi\pip\pim}}}
      {\BR_{\decay{\Bu}{\psitwos\Kp}} \times 
       \BR_{\decay{\psitwos}{\jpsi\pip\pim}}} 
       & = &   \left(3.69 \pm 0.07 \pm 0.06 \right) \times 10^{-2}\,, 
\end{eqnarray*}
where the first uncertainty  is statistical and 
the~second is systematic. 
The~decay of \mbox{\BpTopsiK} 
with \mbox{$\decay{\psitwod}{\jpsi\pip\pim}$}
is observed for the~first time
with a~significance of 5.1~standard
deviations.
The~mass differences between the~\psitwod, \chiconex and \psitwos~states are measured to be
\begin{eqnarray*}
m_{\chiconex} - m_{\psitwod} & = &  \phantom{0}47.50  \pm 0.53 \pm 0.13 \mevcc\,, \\
m_{\psitwod}  - m_{\psitwos} & = &  137.98 \pm 0.53 \pm 0.14 \mevcc\,, \\
m_{\chiconex} - m_{\psitwos} & = &  185.49 \pm 0.06 \pm 0.03 \mevcc\,,
\end{eqnarray*}
resulting in the~most precise determination of 
the~\chiconex~mass. 
The~width of the~\psitwod~state is found to be 
below 5.2\mev  at 90\% confidence level.
The~\mbox{Breit}\nobreakdash--Wigner width
of the~\chiconex~state is measured to be
\begin{equation*}
   \Gamma^{\mathrm{BW}}_{\chiconex} = 0.96^{\,+\,0.19}_{\,-\,0.18}\pm0.21\mev\,,
\end{equation*}
which is inconsistent with zero 
by 5.5~standard deviations. 

\end{abstract}

\vspace*{0.5cm}

\begin{center}
  Published in JHEP 08\,(2020) 123 
\end{center}

\vspace{\fill}

{\footnotesize 
% Edit macro in main.tex to keep metadata correct
\centerline{\copyright~\papercopyright. \href{\paperlicenceurl}{\paperlicence}.}}
\vspace*{2mm}

\end{titlepage}

%%%%%%%%%%%%%%%%%%%%%%%%%%%%%%%%
%%%%%  EOD OF TITLE PAGE  %%%%%%
%%%%%%%%%%%%%%%%%%%%%%%%%%%%%%%%

%  empty page follows the title page ----
\newpage
\setcounter{page}{2}
\mbox{~}
%\newpage
%
%% Author List ----------------------------
%%  You need to get a new author list!
%\input{LHCb_authorlist.tex}

\cleardoublepage

%\twocolumn
% %%%%%%%%%%%%% ---------

\renewcommand{\thefootnote}{\arabic{footnote}}
\setcounter{footnote}{0}

%%%%%%%%%%%%%%%%%%%%%%%%%%%%%%%%
%%%%%  Table of Content   %%%%%%
%%%%%%%%%%%%%%%%%%%%%%%%%%%%%%%%
%%%% Uncomment next 2 lines if desired
%\tableofcontents
%\cleardoublepage

%%%%%%%%%%%%%%%%%%%%%%%%%
%%%%% Main text %%%%%%%%%
%%%%%%%%%%%%%%%%%%%%%%%%%

\pagestyle{plain} % restore page numbers for the main text
\setcounter{page}{1}
\pagenumbering{arabic}

%% Uncomment during review phase. 
%% Comment before a final submission.
%% \linenumbers

% You can include short sections directly in the main tex file.
% However, for larger papers it is desirable to split the text into
% several semiautonomous files, which can be revised independently.
% This is especially useful when developing a document in
% collaboration with several people, since then different parts can be
% edited independently.  This type of file organization is shown here.
% 

\section{Introduction}
\label{sec:Introduction}
The~observation of  a~narrow \chiconex~state 
in the~$\jpsi\pip\pim$~mass spectrum of~\mbox{$\decay{\Bu}{\jpsi\pip\pim\Kp}$} decays 
by the~Belle collaboration in 2003~\cite{Choi:2003ue}
has led to a renewed interest in the~study of hadrons containing heavy quarks.
Many new charmonium\nobreakdash-like 
states have since been observed~\cite{PDG2019}.
Some of the new states are unambiguously interpreted as 
conventional  $\cquark\cquarkbar$~states, 
some are manifestly exotic~\cite{
Choi:2007wga,
Mizuk:2008me,
Mizuk:2009da,
Chilikin:2013tch,
LHCb-PAPER-2014-014,
LHCb-PAPER-2015-038,
LHCb-PAPER-2018-034},
while for the~others a~definite interpretation is still missing~\cite{Olsen:2017bmm,Brambilla:2019esw,Liu:2019zoy}.
Despite the~large amount of 
experimental data~\cite{Acosta:2003zx,
Abazov:2004kp,
Aubert:2004ns,
Abulencia:2005zc,
Aubert:2005zh,
Aubert:2006aj,
Abulencia:2006ma,
Aubert:2007pz,
Aubert:2008gu,
Aaltonen:2009vj,
Aubert:2008ae,
Adachi:2008sua,
delAmoSanchez:2010jr,
Choi:2011fc,
Bhardwaj:2011dj,
LHCb-PAPER-2011-034,
LHCb-PAPER-2013-001,
Chatrchyan:2013cld,
Ablikim:2013dyn,
LHCb-PAPER-2014-008,
Bala:2015wep,
LHCb-PAPER-2015-015,
Aaboud:2016vzw,
LHCb-PAPER-2016-016,
LHCb-PAPER-2019-023,
Chou:2019mlv,
Bhardwaj:2019spn,
Sirunyan:2020qir},
the~nature of the~\chiconex~state is still unclear.
Several interpretations 
have been proposed, 
such as
a~conventional 
$\Pchi_{\cquark1}\mathrm{(2P)}$~state~\cite{Achasov:2015oia},
a~molecular state~\cite{Tornqvist:2004qy,Swanson:2003tb,Wong:2003xk},
a~tetraquark~\cite{Maiani:2004vq},
a~$\cquark\cquarkbar\mathrm{g}$~hybrid state~\cite{Li:2004sta},
a~vector glueball~\cite{Seth:2004zb} or 
a~mixed state~\cite{Matheus:2009vq, Chen:2013pya}.
Precise measurements of 
the~resonance parameters, 
namely the~mass and the~width, 
are crucial for the~correct interpretation of the~state.
Comparison of the~decays of beauty hadrons with 
final states involving the~\chiconex~particle and those 
involving other charmonium resonances can shed light on the~production 
mechanism, in particular, on the~role of 
$\Dz\Dstarzb$~rescattering~\cite{Braaten:2019yua}. 
%%
%% Recently, the~LHCb collaboration performed an~innovative analysis
%% of the~lineshape of the~\chiconex state, 
%% reconstructed via the~\JpsiPiPi~final state, and determined
%% the~resonance parameters 
%% with an~unprecedented precision 
%% and searched for the~poles of 
%% the~complex Flatt\'e\nobreakdash-like 
%% amplitude~\cite{LHCb-PAPER-2020-008}.

%%
A~recent analysis of 
$\Dz\Dzb$ and $\Dp\Dm$~mass spectra, performed  by the~LHCb collaboration~\cite{LHCb-PAPER-2019-005},   
led to the~observation of a~new narrow state, $\Ppsi_3\mathrm{(3842)}$, 
interpreted as~a~spin\nobreakdash-3 component of 
the~D\nobreakdash-wave charmonium triplet,  $\Ppsi_3\mathrm{(1^3D_3)}$~state~\cite{Piemonte:2019cbi,Yu:2019kcr}, 
and a~precise measurement of the~mass of 
the~vector component of this triplet, 
the~$\Ppsi\mathrm{(3770)}$~state.
Evidence for the~third, tensor component of the~triplet,
the~\psitwod state\footnote{A hint for this state was 
reported  in 1994 by the~E705 experiment in studies 
of the~\JpsiPiPi final state in pion\nobreakdash-lithium collisions 
with a~statistical significance of 
$2.8$~standard deviations~\cite{PhysRevD.50.4258}. 
%% The~precision of the~measurement and knowledge of 
%% the~momentum scale means that the~signal could also be 
%% interpreted as being due to the~\chiconex state. 
}, 
%% in the~$\decay{\psitwod}{\chicone\g}$~decay 
was reported by the~\belle collaboration 
in the~$\decay{\B}{\left(\decay{\psitwod}{\Pchi_{\cquark1}\g}\right)\kaon}$~decays~\cite{Bhardwaj:2013rmw}. 
This~was 
confirmed by the~\besiii~collaboration with a~significance in excess of 
5~standard deviations~\cite{PhysRevLett.115.011803}.
The~partial decay widths of the~\psitwod resonance are calculated 
to be 
\mbox{$\Gamma_{\decay{\psitwod}{\chicone\g}}=215\kev$}~\cite{Ebert:2002pp},
\mbox{$\Gamma_{\decay{\psitwod}{\chictwo\g}}=59\kev$}~\cite{Ebert:2002pp},
\mbox{$\Gamma_{\decay{\psitwod}{\mathrm{ggg}}}=36\kev$}~\cite{Eichten:2002qv},
and 
\mbox{$\Gamma_{\decay{\psitwod}{\jpsi\Ppi\Ppi}}\simeq160\kev$}~\cite{Wang:2015xsa}, 
corresponding to a~total width of 470\kev
and 
a~branching fraction 
\mbox{$\BR_{\decay{\psitwod}{\jpsi\Ppi\Ppi}}$}  of  $34\%$~\cite{Xu:2016kbn}.
The~predicted width is much smaller than the~upper 
limit  of 16\mev at 90\%~confidence level\,(CL) set by the~\besiii 
collaboration~\cite{PhysRevLett.115.011803}.

In this paper, a~sample of
%% $\decay{\Bu}{\jpsi\pip\pim\Kp}$
\mbox{$\decay{\Bu}{\left(\decay{\PX_{\cquark\cquarkbar}}{\jpsi\pip\pim}\right)\Kp}$}~decays\footnote{Inclusion of charge\nobreakdash-conjugate states is implied throughout the~paper.} is analysed,
where $\PX_{\cquark\cquarkbar}$ denotes
the~\psitwod,
\chiconex
or \psitwos~state and  the~\jpsi~meson is reconstructed in 
the~$\mumu$~final state. 
The~study is based on proton\nobreakdash-proton\,($\proton\proton$) 
collision data,  corresponding to an~integrated luminosity 
of 1, 2, and 6\invfb, 
collected with the~LHCb detector at centre\nobreakdash-of\nobreakdash-mass 
energies of 7, 8, and 13\tev, respectively.
This data sample allows studies of the~properties of 
the~\psitwod and \chiconex ~states produced in 
\B~decay recoiling against a~kaon.  
The~presence of the~\psitwos state in the~same sample 
provides a~convenient sample for normalisation and 
reduction of potential systematic uncertainties.
%%%%
A~complementary measurement using 
inclusive $\decay{\bquark}{\left(\decay{\chiconex}{\jpsi\pip\pim} \right)\PX}$~decays 
and a~data set, corresponding to an~integrated luminosity of 1 and 2\invfb,
collected at the~centre\nobreakdash-of\nobreakdash-mass energies of 7 and 8\tev, 
is reported in Ref.~\cite{LHCb-PAPER-2020-008}.
This~gives a~determination of 
the~resonance parameters for the~\chiconex~state
with an~unprecedented precision, including
searches for the~poles of 
the~complex Flatt\'e\nobreakdash-like 
amplitude.

\section{Detector and simulation}
\label{sec:Detector}

The \lhcb detector~\cite{Alves:2008zz,LHCb-DP-2014-002} is a single-arm forward
spectrometer covering the~\mbox{pseudorapidity} range $2<\eta <5$,
designed for the study of particles containing $\bquark$~or~$\cquark$~quarks. 
The~detector includes a high-precision tracking system consisting of a 
silicon-strip vertex detector surrounding the \proton\proton interaction
region~\cite{LHCb-DP-2014-001}, a large-area silicon-strip detector located
upstream of a dipole magnet with a bending power of about $4{\mathrm{\,Tm}}$,
and three stations of silicon-strip detectors and straw drift tubes~\cite{LHCb-DP-2013-003,LHCb-DP-2017-001} placed downstream of the magnet. 
The tracking system provides a measurement of the momentum of charged particles
with a relative uncertainty that varies from $0.5\%$ at low momentum to $1.0\%$~at~$200 \gevc$. 
The~momentum scale is calibrated using samples of $\decay{\jpsi}{\mumu}$ 
and $\decay{\Bu}{\jpsi\Kp}$~decays collected concurrently
with the~data sample used for this analysis~\cite{LHCb-PAPER-2012-048,LHCb-PAPER-2013-011}. 
The~relative accuracy of this
procedure is estimated to be $3 \times 10^{-4}$ using samples of other
fully reconstructed $\bquark$~hadrons, $\PUpsilon$~and
$\KS$~mesons.
The~minimum distance of a track to a primary $\proton\proton$\nobreakdash-collision vertex\,(PV), 
the~impact parameter\,(IP), 
is~measured with a~resolution of $(15+29/\pt)\mum$, where \pt is the component 
of the~momentum transverse to the beam, in\,\gevc. Different types of charged hadrons
are distinguished using information from two ring-imaging Cherenkov detectors\,(RICH)~\cite{LHCb-DP-2012-003}. Photons,~electrons and hadrons are identified 
by a~calorimeter system consisting of scintillating\nobreakdash-pad 
and preshower detectors, 
an electromagnetic and 
a~hadronic calorimeter. Muons are~identified by a~system 
composed of alternating layers of iron and multiwire proportional chambers~\cite{LHCb-DP-2012-002}.

The online event selection is performed by a trigger~\cite{LHCb-DP-2012-004}, 
which consists of a hardware stage, based on information from the calorimeter and muon systems,
followed by a~software stage, which applies a~full event reconstruction. 
The hardware trigger selects muon candidates with high transverse momentum or dimuon candidates with a~high value of 
the~product
of the~$\pt$ of each muon. 
In~the~software trigger two 
oppositely charged muons are required to form 
a~good\nobreakdash-quality
vertex that is significantly displaced from every~PV,
with a~dimuon mass exceeding~$2.7\gevcc$.

Simulated events are used to describe the~signal  
shapes
and to~compute efficiencies, needed to determine 
the~branching fraction ratios.
In~the~simulation, \proton\proton collisions are generated 
using \pythia~\cite{Sjostrand:2007gs}  with a~specific \lhcb configuration~\cite{LHCb-PROC-2010-056}. 
Decays of unstable particles are described by 
the~\evtgen 
package~\cite{Lange:2001uf}, 
in which final-state radiation is generated using \photos~\cite{Golonka:2005pn}. 
The \PsitwodTojpsipipi decays are simulated using a phase\nobreakdash-space model. 
The~\ChiconexTojpsipipi decays are simulated  proceeding
via the~S\nobreakdash-wave $\jpsi\Prho^0$ 
intermediate state~\cite{LHCb-PAPER-2015-015}. 
For the~\psitwos decays 
the~model described in 
Refs.~\cite{Gottfried:1977gp,Voloshin:1978hc,Peskin:1979va,Bhanot:1979vb} is used.
The~interaction of the~generated particles with the~detector, 
and its response, are implemented using
the~\geant toolkit~\cite{Allison:2006ve, *Agostinelli:2002hh} 
as described in Ref.~\cite{LHCb-PROC-2011-006}.
To~account for imperfections in the~simulation of
charged\nobreakdash-particle reconstruction, 
the~track reconstruction efficiency
determined from simulation 
is corrected using data-driven techniques~\cite{LHCb-DP-2013-002}.

\section{Event selection}
\label{sec:Selection}
 
 Candidate \BpToJpsipipiK~decays  are reconstructed using the
 \decay{\jpsi}{\mumu} decay mode. 
%%To~separate signal from background, 
A~loose preselection
 similar to Refs.~\cite{
 LHCb-PAPER-2012-010,
 LHCb-PAPER-2013-010,
 LHCb-PAPER-2013-024,
 LHCb-PAPER-2013-047,
 LHCb-PAPER-2014-009,
 LHCb-PAPER-2014-060,
 LHCb-PAPER-2015-060,
 LHCb-PAPER-2016-040,
 LHCb-PAPER-2018-022,
 LHCb-PAPER-2019-023,
 LHCb-PAPER-2019-025,
 LHCb-PAPER-2019-045} is applied, 
 followed by a~multivariate classifier based on 
 a~decision tree with gradient boosting\,(BDT)~\cite{BDTG}. 
 
 Muon, pion and kaon candidates are identified 
 by combining information from the~RICH, 
 calorimeter and muon detectors~\cite{LHCb-PROC-2011-008}. 
The~transverse momentum of muon\,(hadron)
candiates is required to be larger than 550\,(220)\mevc.
 To~allow for efficient  particle identification,
 kaons and pions are required to have a~momentum between 3.2~and~150\gevc. 
 To~reduce combinatorial background, 
 only tracks that are inconsistent with originating from any 
 reconstructed PV in the~event are considered.
Pairs of oppositely charged muons consistent with originating from a~common vertex 
are combined to form \decay{\jpsi}{\mumu} candidates. 
The~reconstructed mass of the~pair is required to be
between  $3.0$ and $3.2 \gevcc$. 
 
 To form the $\Bp$~candidates, 
 the selected $\jpsi$~candidates 
 are combined with a pair of oppositely charged pions
and a~positively charged kaon.  Each~$\Bp$~candidate is associated with
 the~PV that yields the~smallest~$\chisqip$, 
 where \chisqip is defined as the~difference in the~vertex\nobreakdash-fit 
 \chisq of a~given PV 
 reconstructed with and without the~particle under consideration. 
 %% The~$\chisqip$ value is required to be less than~9. 
 To~improve
 the~mass resolution for the~$\Bp$~candidates,
 a~kinematic fit~\cite{Hulsbergen:2005pu}  is performed. 
 This~fit constrains the~mass of the~$\mumu$~pair to the~known mass
 of the~$\jpsi$~meson~\cite{PDG2019} and  
 constraints the~\Bu~candidate 
 to originate from its associated~PV.
 In~addition, the~measured decay time of 
 the~$\Bp$~candidate, 
 calculated with respect to the~associated PV, 
 is required to be greater than $75\mum/c$. 
 This~requirement suppresses
background from particles originating 
 from the~PV.

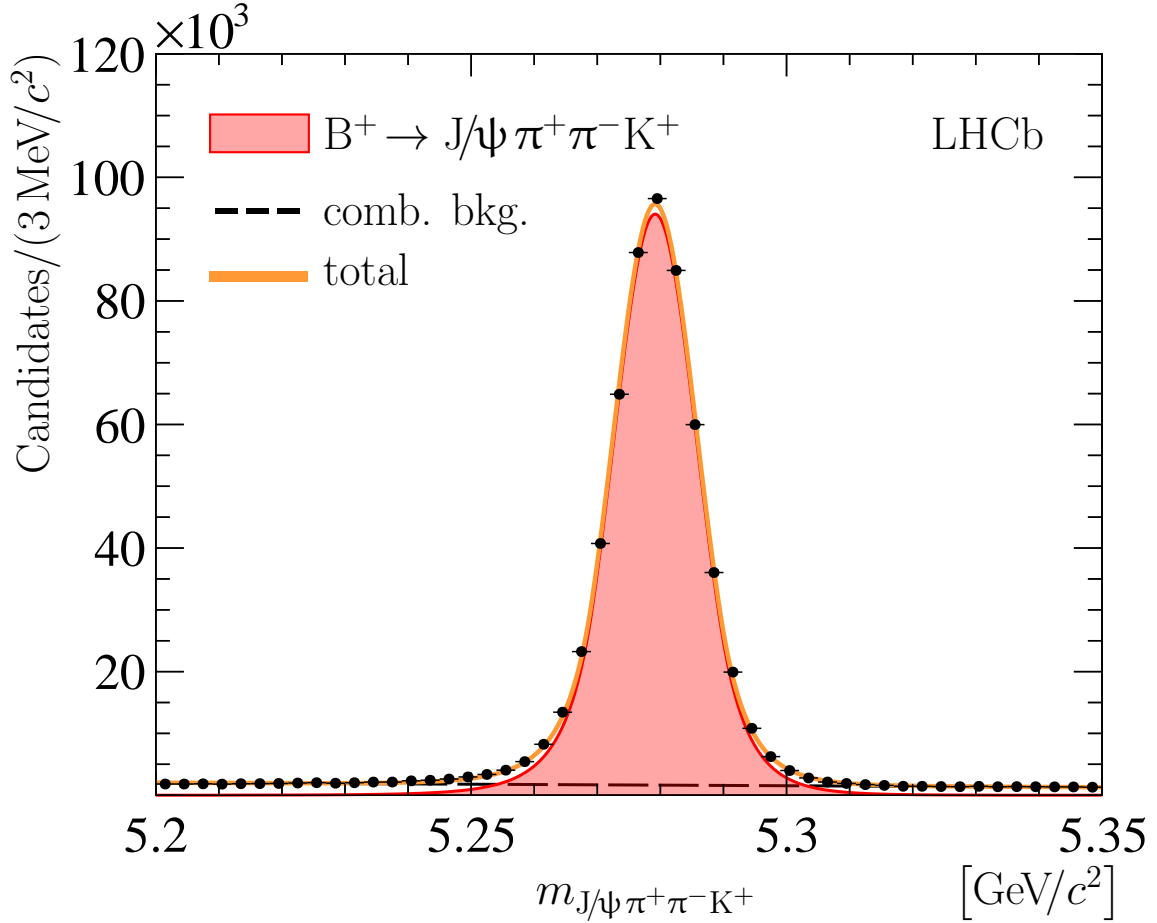
\begin{figure}[t]
  \setlength{\unitlength}{1mm}
  \centering
  \begin{picture}(150,125)
    %%\graphpaper[5](-10,-10)(170,145)
    \put(0,3){\includegraphics*[width=150mm,height=120mm]{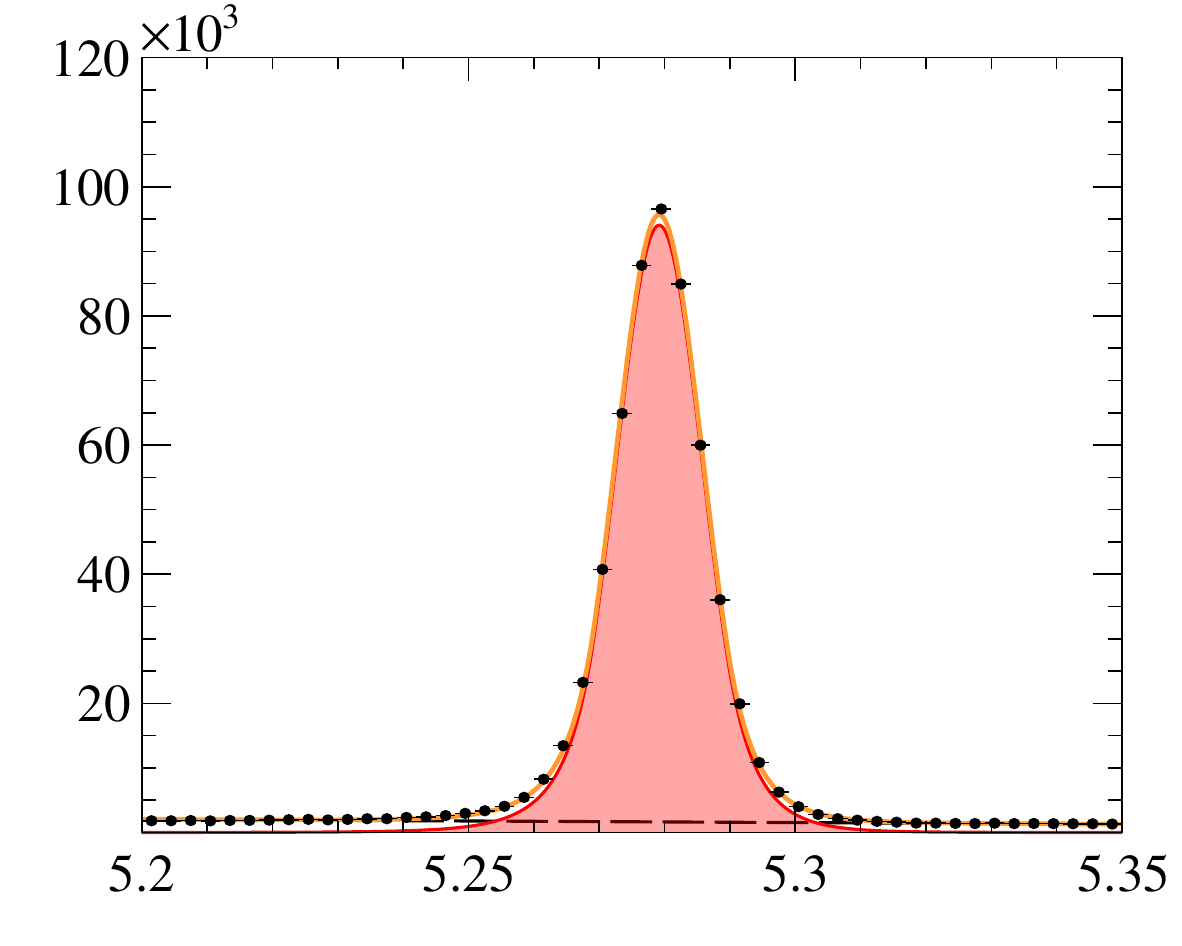}}
    \put( -1,60){\begin{sideways}\Large{Candidates/$(3\mevcc)$}\end{sideways}}
    \put( 68, 3){\Large$m_{\jpsi\pip\pim\Kp}$}
    \put(123, 3){\Large$\left[\!\gevcc\right]$}
    \put(120,103){\Large\lhcb}
    \put(25,102.5) {\begin{tikzpicture}[x=1mm,y=1mm]\filldraw[fill=red!35!white,draw=red,thick]  (0,0) rectangle (13,5);\end{tikzpicture} }
    \put(25,94.5){\color[RGB]{0,0,0}     {\hdashrule[0.0ex][x]{13mm}{2.0pt}{3.0mm 0.3mm} } }
    \put(25,85.5){\color[RGB]{255,153,51} {\rule{13mm}{4.0pt}}}
    \put( 40,103){\Large{\decay{\Bp}{\jpsi\pip\pim\Kp}}}
    \put( 40,93){\Large{comb. bkg.}}
    \put( 40,85){\Large{total}}
  \end{picture}
  \caption {\small 
    Distribution  for 
    the~\mbox{$\jpsi\pip\pim\Km$}~mass
    for selected \Bu~candidates\,(points with error bars). 
    A~fit, described in the~text,  is overlaid.}
  \label{fig:signalalt}
\end{figure}

 A BDT is used to further suppress the combinatorial 
 background. 
 It~is~trained using a~simulated sample of 
 \mbox{$\decay{\Bu}{\left( \decay{\psitwod}{\jpsi\pip\pim}\right)\Kp}$}~decays
as the signal. For~the~background, a~sample of 
 $\jpsi\pip\pip\Km$~combinations with same\nobreakdash-sign pions in~data, 
 passing the~preselection criteria and having the~mass in 
 the~range between 5.20 and 5.35\gevcc, is used.
 The~$k$\nobreakdash-fold cross-validation 
 technique~\cite{geisser1993predictive}
 with $k=13$ is used to avoid introducing
 a~bias in the~BDT~evaluation.
 The~BDT is trained on variables related to
 the~reconstruction quality,
 decay kinematics,
decay time
 of $\Bp$~candidate and  
the~quality of the~kinematic fit.
 The~requirement on the~BDT output is 
 chosen to maximize 
 \mbox{$\upvarepsilon/(\alpha/2 +
 \sqrt{B})$}~\cite{Punzi:2003bu}, 
 where $\upvarepsilon$ is the~signal efficiency 
 for the~\mbox{\BpTopsiK} decays
 obtained from simulation; $\alpha = 5$ 
 is the~target signal significance
 in~units of standard deviations;  
 $B$~is the~expected background 
 yield  within narrow mass windows
 centred at the~known 
 $\Bp$~and $\psitwod$~masses~\cite{PDG2019}.  
 The~mass distribution of selected 
 \mbox{$\decay{\Bu}{\jpsi\pip\pim\Kp}$}~candidates
 is shown in Fig.~\ref{fig:signalalt}.
 The~data are fit with a~sum 
 of a~modified Gaussian function
with power-law tails on both sides~\cite{LHCb-PAPER-2011-013,Skwarnicki:1986xj} 
and a~linear polynomial 
combinatorial background component.
The~\Bu~signal yield 
is \mbox{$(547.8\pm0.8)\times10^3$}~candidates.

\section{Signal yields, masses and widths}
\label{sec:Sig_eff}
The yields for the~\mbox{$\decay{\Bu}{\left(\decay{\PX_{\cquark\cquarkbar}}{\jpsi\pip\pim}\right)\Kp}$}~decays 
%% where $\PX_{\cquark\cquarkbar}$ denotes
%% the~\psitwod,
%% \chiconex
%% or \psitwos~state, 
are determined 
using
%%a~{\color{red}{\sout{simultaneous}}}
a~two\nobreakdash-dimensional
unbinned extended
maximum\nobreakdash-likelihood fit
to the~\mbox{$\jpsi\pip\pim\Kp$}~mass, 
$m_{\jpsi\pip\pim\Kp}$,
and the~\mbox{$\jpsi\pip\pim$}~mass,
$m_{\jpsi\pip\pim}$, distributions.
The~fit is performed
simultaneously
in the~three non\nobreakdash-overlapping regions
%%
%% {\color{blue}{to two\nobreakdash-dimensional
%% (the~\mbox{$\jpsi\pip\pim\Kp$}~mass, 
%% $m_{\jpsi\pip\pim\Kp}$, versus
%% the~\mbox{$\jpsi\pip\pim$}~mass,
%% $m_{\jpsi\pip\pim}$) distributions
%% in the~three non\nobreakdash-overlapping regions}}
\begin{itemize}
   \item $3.67 \le m_{\jpsi\pip\pim}<3.70\gevcc$\,, 
   \item $3.80 \le m_{\jpsi\pip\pim}<3.85\gevcc$\,, 
   \item $3.85 \le m_{\jpsi\pip\pim}<3.90\gevcc$\,, 
\end{itemize}
corresponding to 
the~\mbox{$\decay{\Bu}{\psitwos\Kp}$},
\mbox{$\decay{\Bu}{\psitwod\Kp}$} and 
\mbox{$\decay{\Bu}{\chiconex\Kp}$}~decays.
For~each of the~three regions 
the~\mbox{$\jpsi\pip\pim\Kp$}~mass
is restricted to 
\mbox{$5.20 \le m_{\jpsi\pip\pim\Kp}<5.35\gevcc$}.
To~improve the~resolution on the~$\jpsi\pip\pim$~mass
and to eliminate a~small correlation between 
$m_{\jpsi\pip\pim\Kp}$ and $m_{\JpsiPiPi}$~variables, 
the~$m_{\JpsiPiPi}$ variable is computed
using a~kinematic fit~\cite{Hulsbergen:2005pu} 
that constrains the~mass of the~\Bu~candidate 
to its known value~\cite{PDG2019}.
In~each region, the~fit function is defined as a~sum of four components: 
\begin{enumerate}
\item signal $\decay{\Bp}{\PX_{\cquark\cquarkbar}\Kp}$ decays
  parameterised as a~product of  the~\Bu~and $\PX_{\cquark\cquarkbar}$~signal templates
  described in detail in the~next paragraph; 
\item contribution 
  from 
  the~decays \mbox{\decay{\Bp}{\left(\jpsi\pip\pim\right)_{\mathrm{NR}}\Kp}}
  with no narrow intermediate $\PX_{\cquark\cquarkbar}$~state,
  parameterised as a~product of the~\Bu~signal template and 
  a~linear function of $m_{\jpsi\pip\pim}$; 
\item random combinations of  
  $\PX_{\cquark\cquarkbar}$ and \Kp~candidates,
  parameterised as a~product of the~$\PX_{\cquark\cquarkbar}$~signal template
  and a~linear function of $m_{\jpsi\pip\pim\Kp}$;
\item random $\jpsi\pip\pim\Kp$ combinations, described below.
\end{enumerate}
The~templates for the~\Bp signals 
are described by a~modified Gaussian function 
with power\nobreakdash-law tails on both sides of
the~distribution~\cite{LHCb-PAPER-2011-013,Skwarnicki:1986xj}. 
The~tail parameters are fixed to 
the~values obtained from simulation.
The~narrow $\PX_{\cquark\cquarkbar}$~signal templates are 
parameterised with S\nobreakdash-wave relativistic Breit--Wigner functions 
convolved with the~mass resolution.
Due~to the~proximity of 
the~\chiconex state to the~$\Dz\Dstarzb$~threshold,
modelling this component as a~Breit--Wigner function may not be adequate~\cite{Hanhart:2007yq,
Stapleton:2009ey,
Kalashnikova:2009gt,
Artoisenet:2010va,
Hanhart:2011jz}. 
However, the~analysis from Ref.~\cite{LHCb-PAPER-2020-008} 
demonstrates that a~good description of data 
is obtained with a~Breit\nobreakdash--Wigner 
lineshape when the~mass resolution is included. 
The~mass resolution is described 
by a~symmetric modified Gaussian function 
with power\nobreakdash-law tails 
on both sides of the distribution, 
with the~parameters 
fixed to the~values from simulation.
In~the~template for the~\Bu~signal, the~peak\nobreakdash-position parameter 
is shared between 
all three decays and allowed to vary in the~fit.
The~mass
resolutions used in the~\Bp and $\PX_{\cquark\cquarkbar}$~signal 
templates are fixed to the~values determined from simulation, 
but are corrected by common  scale factors,
$f_{\Bu}$ and $f_{\PX_{\cquark\cquarkbar}}$, 
to account for a~small discrepancy  
in the~mass
resolution  between data and simulation.  
The~masses of the~$\PX_{\cquark\cquarkbar}$~signal templates,
as well as the~Breit--Wigner widths for the~\psitwod and \chiconex~states,
are free fit parameters, 
while the~width in
the~template for the~\psitwos~signal is fixed to its~known value~\cite{PDG2019}.
The~combinatorial\nobreakdash-background component is modelled
with a~smooth two\nobreakdash-dimensional function
\begin{equation}
  %% f(m_{\jpsi\pip\pim\Kp}, 
  %% m_{\jpsi\pip\pim}) & = 
  {\mathcal{E}}(m_{\jpsi\pip\pim\Kp})
  \times   
      {\mathcal{P}}_{3,4}(m_{\jpsi\pip\pim}) 
      \times  
      P_{\mathrm{2D}}(m_{\jpsi\pip\pim\Kp}, 
      m_{\jpsi\pip\pim}),
\end{equation}
where
${\mathcal{E}}(m_{\jpsi\pip\pim\Kp})$ is an~exponential function,
${\mathcal{P}}_{3,4}(m_{\jpsi\pip\pim})$ is 
%% a~three\nobreakdash-body\,(\jpsi\pip\pim)
%% phase\nobreakdash-space function of 
%% the~four\nobreakdash-body \Bp~decay~\cite{Byckling},
a~three\nobreakdash-body 
phase\nobreakdash-space function~\cite{Byckling},
and~$P_{\mathrm{2D}}$ is a~two\nobreakdash-dimensional positive 
bilinear function, 
which accounts for small non\nobreakdash-factorizable effects.  
For~the~considered fit ranges 
${\mathcal{P}}_{3,4}(m_{\jpsi\pip\pim})$ is
close to a~constant. 
%% practically a~constant function.
%%% The~later accounts for possible correlation

\begin{figure}[t]
  \setlength{\unitlength}{1mm}
  \centering
  \begin{picture}(150,180)
    \definecolor{root8}{rgb}{0.35, 0.83, 0.33}
    %\graphpaper[5](-10,-10)(170,140)
    
    \put( 0,120){\includegraphics*[width=75mm,height=60mm]{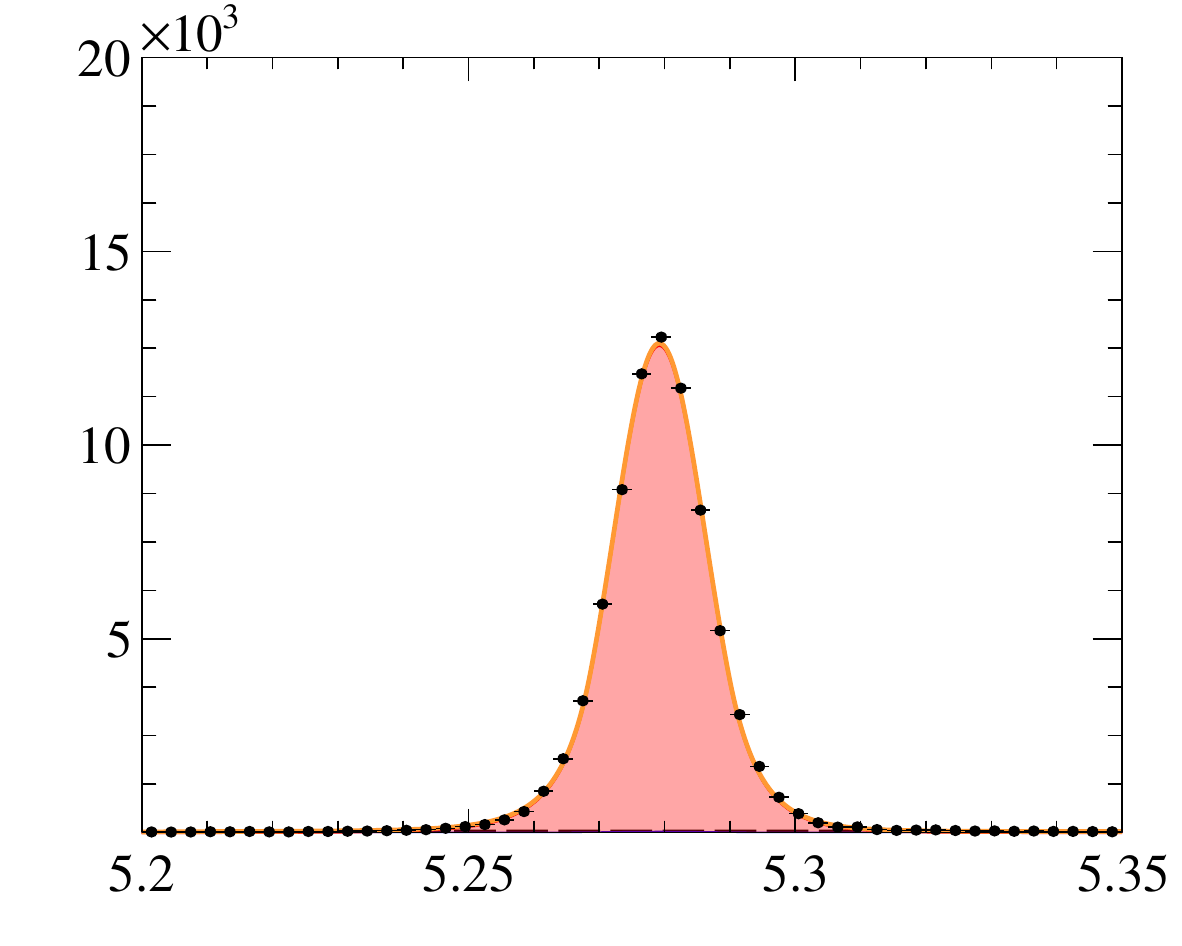}}
    \put(80,120){\includegraphics*[width=75mm,height=60mm]{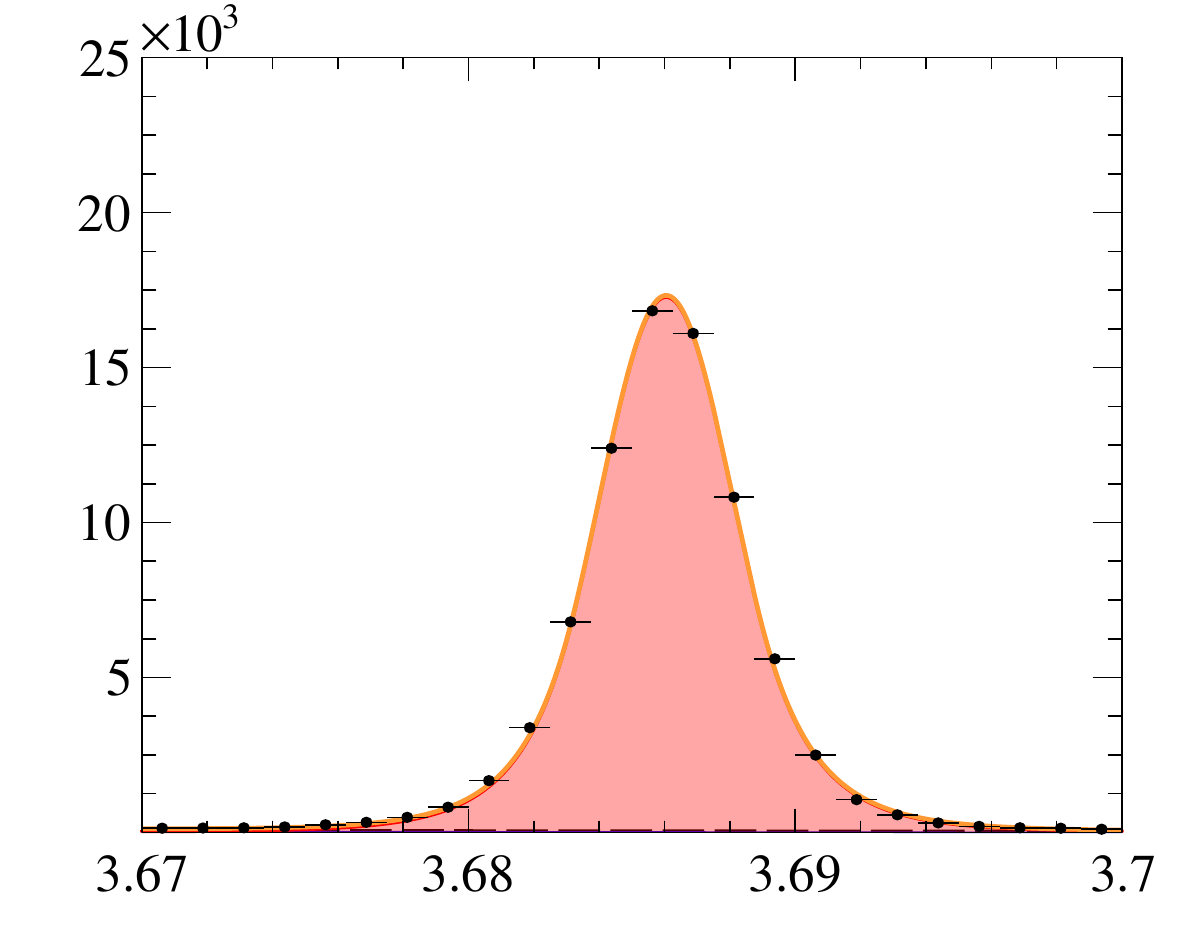}}
    \put( 0, 60){\includegraphics*[width=75mm,height=60mm]{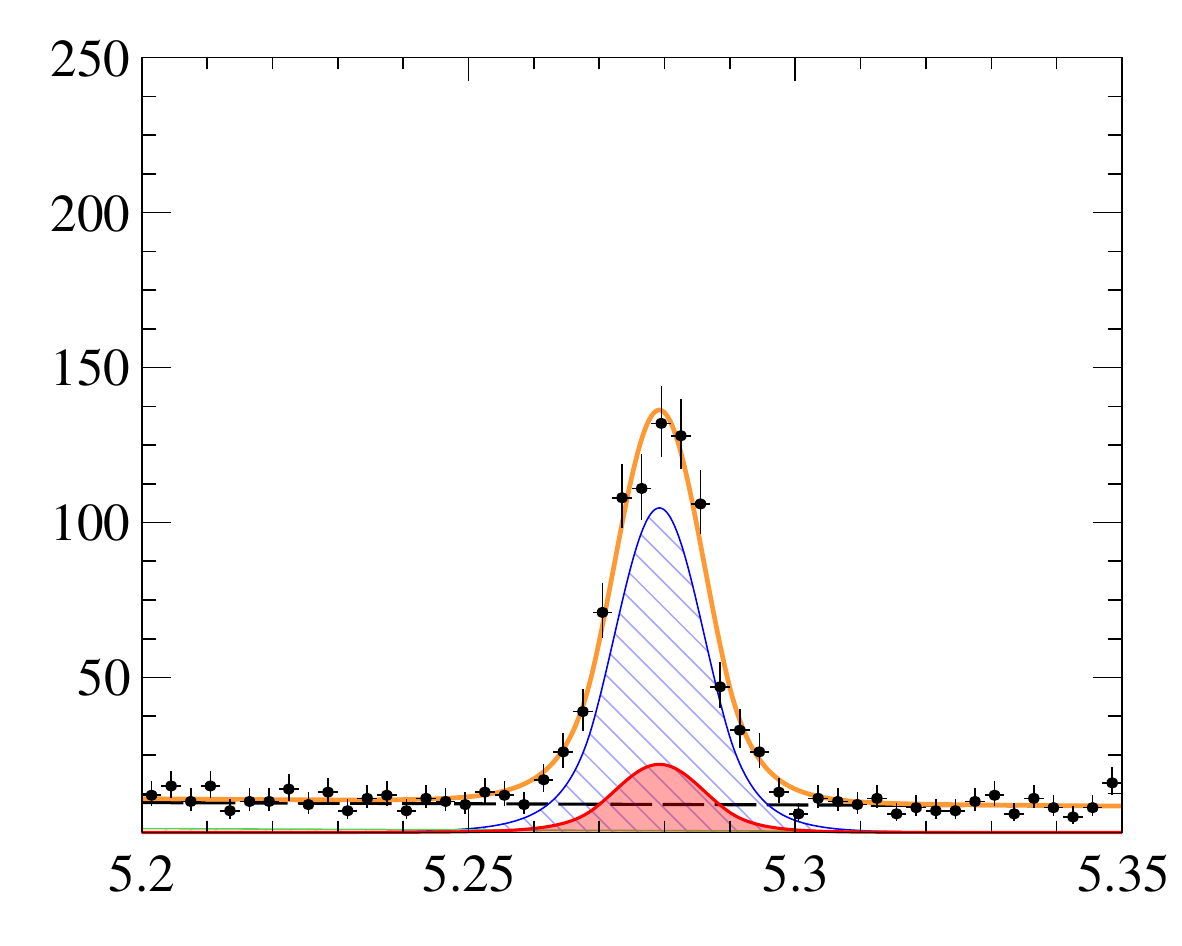}}
    \put(80, 60){\includegraphics*[width=75mm,height=60mm]{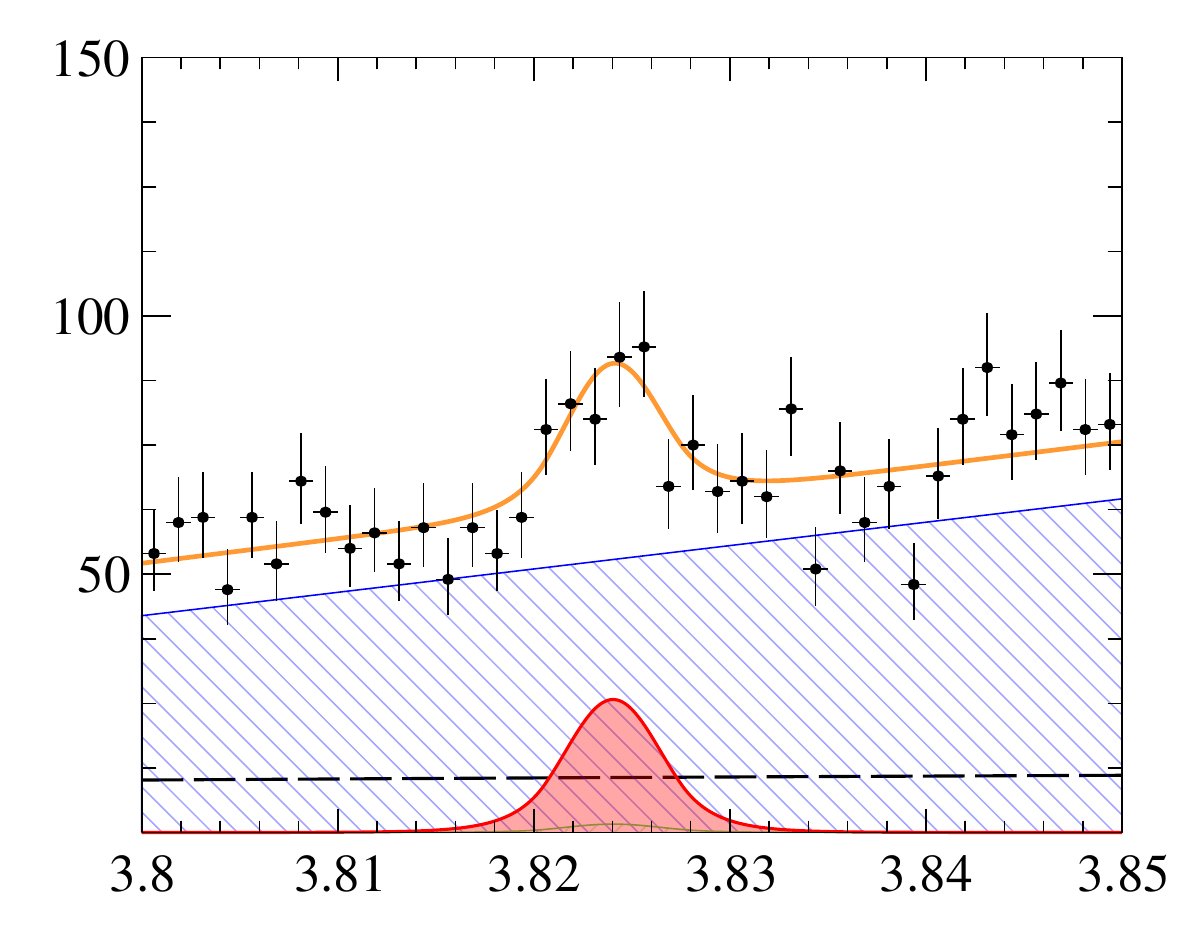}}
    \put( 0,  0){\includegraphics*[width=75mm,height=60mm]{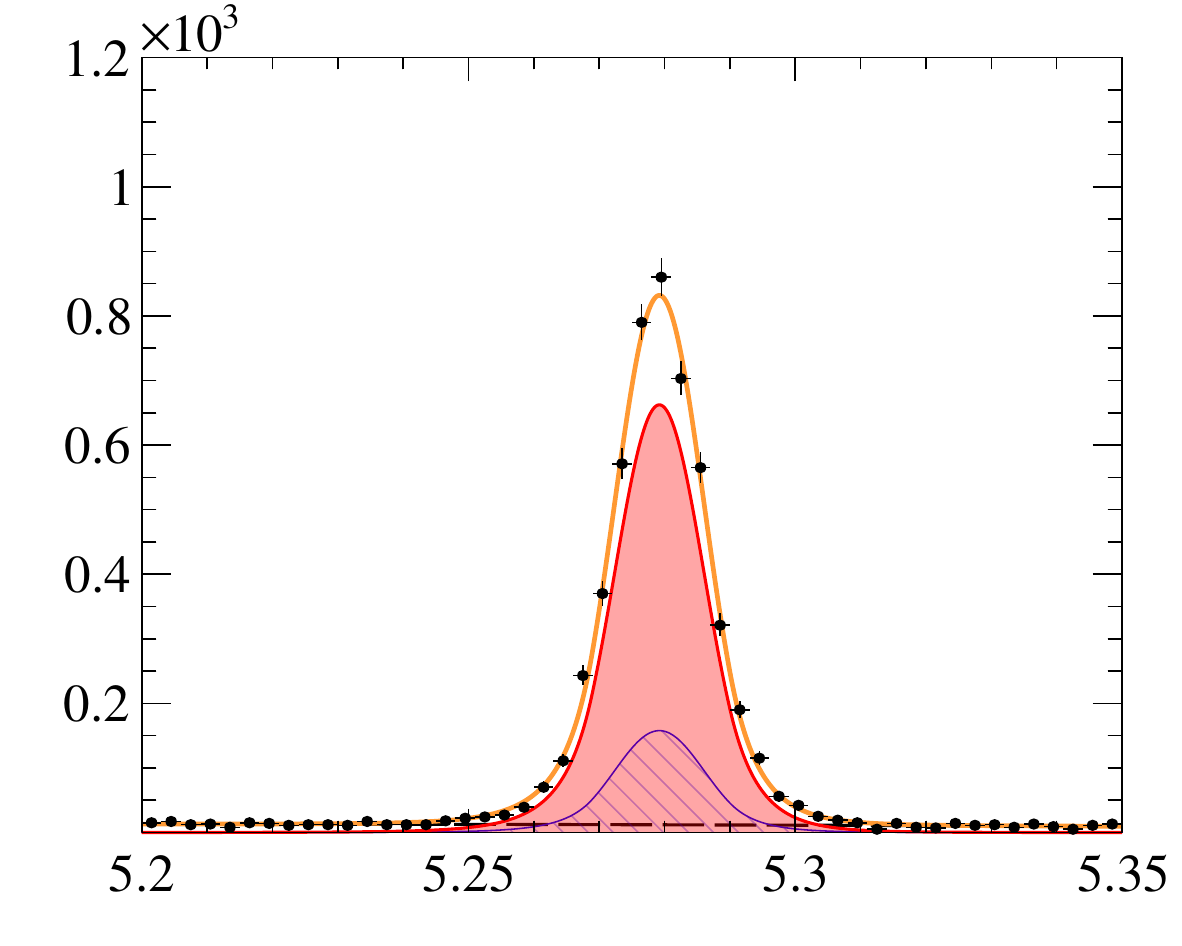}}
    \put(80,  0){\includegraphics*[width=75mm,height=60mm]{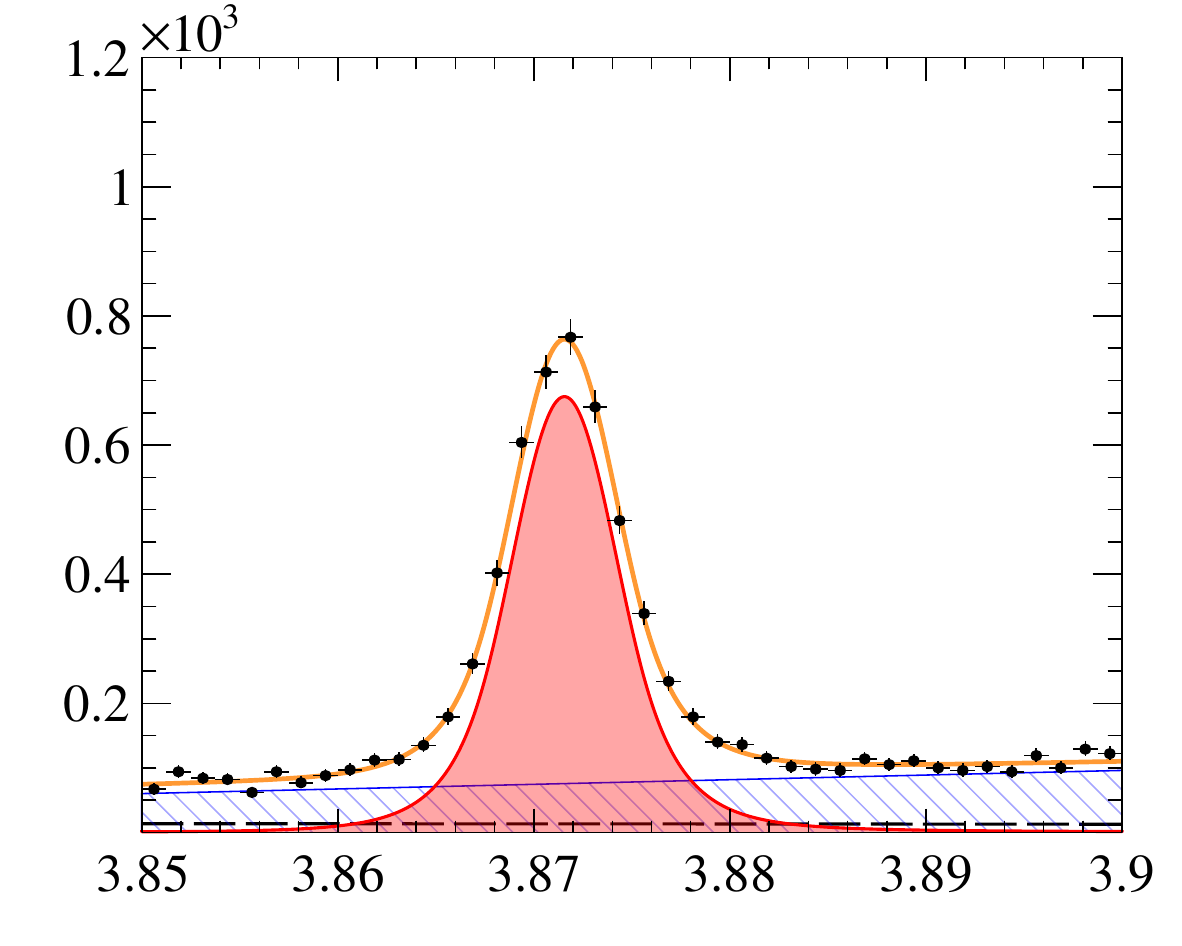}}
    
    \put(  -3,138){\begin{sideways}\small{Candidates/$(3\mevcc)$}\end{sideways}}
    \put(  -3,78){\begin{sideways}\small{Candidates/$(3\mevcc)$}\end{sideways}}
    \put(  -3,18){\begin{sideways}\small{Candidates/$(3\mevcc)$}\end{sideways}}
    
    \put(  78,133){\begin{sideways}\small{Candidates/$(1.25\mevcc)$}\end{sideways}}
    \put(  78,73){\begin{sideways}\small{Candidates/$(1.25\mevcc)$}\end{sideways}}
    \put(  78,13){\begin{sideways}\small{Candidates/$(1.25\mevcc)$}\end{sideways}}
    
    \put(30 ,120){$m_{\jpsi\pip\pim\Kp}$}
    \put(30 ,60){$m_{\jpsi\pip\pim\Kp}$}
    \put(30 ,0){$m_{\jpsi\pip\pim\Kp}$}
    
    \put(113 ,120){$m_{\jpsi\pip\pim}$}
    \put(113 ,60){$m_{\jpsi\pip\pim}$}
    \put(112 ,0){$m_{\jpsi\pip\pim}$}
    
    \put( 57,119){$\left[\!\gevcc\right]$}
    \put( 57,59){$\left[\!\gevcc\right]$}
    \put( 57,-1){$\left[\!\gevcc\right]$}
    
    \put( 137,119){$\left[\!\gevcc\right]$}
    \put( 137,59){$\left[\!\gevcc\right]$}
    \put( 137,-1){$\left[\!\gevcc\right]$}
    
    \put( 12,171){\scriptsize$3.68<m_{\jpsi\pip\pim}<3.69\gevcc$}
    \put( 91,171){\scriptsize$5.26<m_{\jpsi\pip\pim\Kp}<5.30\gevcc$}
    \put( 12,111){\scriptsize$3.82<m_{\jpsi\pip\pim}<3.83\gevcc$}
    \put( 91,111){\scriptsize$5.26<m_{\jpsi\pip\pim\Kp}<5.30\gevcc$}
    \put( 12,51){\scriptsize$3.86<m_{\jpsi\pip\pim}<3.88\gevcc$}
    \put( 91,51){\scriptsize$5.26<m_{\jpsi\pip\pim\Kp}<5.30\gevcc$}
    
    \put( 58,171){\small\lhcb}
    \put( 58,111){\small\lhcb}
    \put( 58,51){\small\lhcb}
    \put( 139,171){\small\lhcb}
    \put( 139,111){\small\lhcb}
    \put( 139,51){\small\lhcb}
    \put(12,105.5) {\begin{tikzpicture}[x=1mm,y=1mm]\filldraw[fill=red!35!white,draw=red,thick]  (0,0) rectangle (8,3);\end{tikzpicture} }
    \put(12,100.5){\begin{tikzpicture}[x=1mm,y=1mm]\draw[thin,blue,pattern=north west lines, pattern color=blue]  (0,0) rectangle (8,3);\end{tikzpicture} }
    \put(12,95.5){\begin{tikzpicture}[x=1mm,y=1mm]\draw[thin,root8,pattern=north east lines, pattern color=root8]  (0,0) rectangle (8,3);\end{tikzpicture} }
    \put(12,92.5){\color[RGB]{0,0,0}     {\hdashrule[0.0ex][x]{8mm}{1.0pt}{2.0mm 0.3mm} } }
    \put(12,88.5){\color[RGB]{255,153,51} {\rule{8mm}{2.0pt}}}
    
    \put( 22,106.3){\scriptsize{\decay{\Bp}{\PX_{\cquark\bar{\cquark}}\Kp}}}
    \put( 22,101){\scriptsize{\decay{\Bp}{\left(\jpsi\pip\pim\right)_{\mathrm{NR}}\Kp}}}
    \put( 22,96.5){\scriptsize{comb. $\PX_{\cquark\bar{\cquark}}\Kp$}}
    \put( 22,92){\scriptsize{comb. bkg.}}
    \put( 22,88){\scriptsize{total}}
  \end{picture}
  \caption {\small 
    Distributions of 
    the~(left)\,\jpsi\pip\pim\Kp and 
    (right)\,\JpsiPiPi mass
    for selected (top)\,\BpTopsitwosK, 
    (middle)\,\BpTopsiK and 
    (bottom)\,\mbox{\BpTochicK}~candidates shown as points with error bars.
    A~fit, described in the~text, is overlaid.}
  \label{fig:signal_fit}
\end{figure}

The~\jpsi\pip\pim\Kp and \JpsiPiPi mass distributions 
together with projections of the~simultaneous unbinned
maximum\nobreakdash-likelihood  fit 
are shown in Fig.~\ref{fig:signal_fit}.
Signal yields $N_{\decay{\Bu}{\PX_{\ccbar}\Kp}}$,  
calculated
mass differences \mbox{$\delta m_{\PX_{\cquark\cquarkbar}}\equiv
m_{\PX_{\cquark\cquarkbar}} - m_{\psitwos}$},
Breit--Wigner widths  $\Gamma_{\PX_{\cquark\cquarkbar}}$
and resolution scale factors 
%% for \Bu~and 
%% $\PX_{\ccbar}$~signal
%% templates, 
%% $f_{\Bu}$ and $f_{\PX_{\cquark\cquarkbar}}$
are listed in 
Table~\ref{tab:sim_fit_res_check}.
%%
%%{\color{blue}{
    The~fit model
    is tested using pseudoexperiments
    and no bias is found in the~results
    and their associated uncertainties.
%%}}
%%
%%{\color{blue}{
    The~masses of \Bp and \psitwos mesons
    are found to be compatible with their known values~\cite{PDG2019}.
%%}}
%%
The~fit component corresponding to 
the~\mbox{$\decay{\Bu}{\left(\jpsi\pip\pim\right)_{\mathrm{NR}}\Kp}$}  
is found to be negligible for 
the~\psitwos~region, dominant for 
the~\psitwod~region  and small for the~\chiconex~region.
The~fit component corresponding
to the~random $\PX_{\ccbar}\Kp$~combinations
is negligible for all fit regions.
The~statistical significance of the~observed 
\mbox{$\decay{\Bu}{\left(\decay{\psitwod}{\jpsi\pip\pim}\right)\Kp}$}
signal over
the~background\nobreakdash-only hypothesis
is estimated to be $5.1$~standard deviations
using Wilks' theorem~\cite{Wilks:1938dza}. 
The~significance is  confirmed by simulating 
a~large number of pseudoexperiments 
according to the~background distribution observed in data.

\begin{table}[b]
  \centering
  \caption{\small 
    Parameters of interest and derived quantities
    from the~simultaneous  unbinned extended 
    maximum-likelihood two-dimensional fit.
    Results and statistical 
    uncertainties are shown for the~three fit regions. 
    %% and described in the~text.
  }
  \vspace{2mm}
  \begin{tabular*}{0.95\textwidth}{@{\hspace{3mm}}l@{\extracolsep{\fill}}lccc@{\hspace{2mm}}}
    \multicolumn{2}{l}{Parameter} & \BpTopsitwosK & \BpTopsiK &   \BpTochicK  
    \\[1.5mm]
    \hline 
    \\[-3mm]
    \multicolumn{2}{l}{$N_{\decay{\Bu}{\PX_{\ccbar}\Kp}}$}
    &  $(81.14\pm0.29)\times10^3$    
    &  $\phantom{0.0}137\pm26\phantom{.0}$                    
    &  $4230 \pm 70$                
    \\
    $\delta m_{\PX_{\ccbar}}$ & $\left[\!\mevcc\right]$ 
    &  --- 
    &  $137.98 \pm 0.53$ 
    &  $185.49 \pm 0.06$
    \\ 
    $\Gamma_{\PX_{\cquark\cquarkbar}}$  & $\left[\!\mev\right]$ 
    &   $0.29$\,(fixed)  %% $0.294$\,(fix)
    &   $\phantom{00.0}0^{\,\,\,+\,\,0.68}_{\,\,\,-\,\,0.00}$
    &   $\phantom{0}0.96^{\,\,\,+\,\,0.19}_{\,\,\,-\,\,0.18}$
    \\[1.5mm]
    \hline 
    \\[-3mm]
    %% $m_{\Bu}$  & $\left[\!\mevcc\right]$ 
    %% &  \multicolumn{3}{c}{ $5279.20 \pm 0.030$ }
    $f_{\Bu}$   
    & \multicolumn{4}{c}{ $1.052\pm0.003$} 
    \\
    $f_{\PX_{\cquark\cquarkbar}}$   
    & \multicolumn{4}{c}{ $1.048\pm0.004$} 
  \end{tabular*}
  \label{tab:sim_fit_res_check}
\end{table}

\begin{figure}[t]
  \setlength{\unitlength}{1mm}
  \centering
  \begin{picture}(150,60)
    %% \graphpaper[5](-10,-10)(170,80)
    
    \put( 0,0){\includegraphics*[width=75mm,height=60mm]{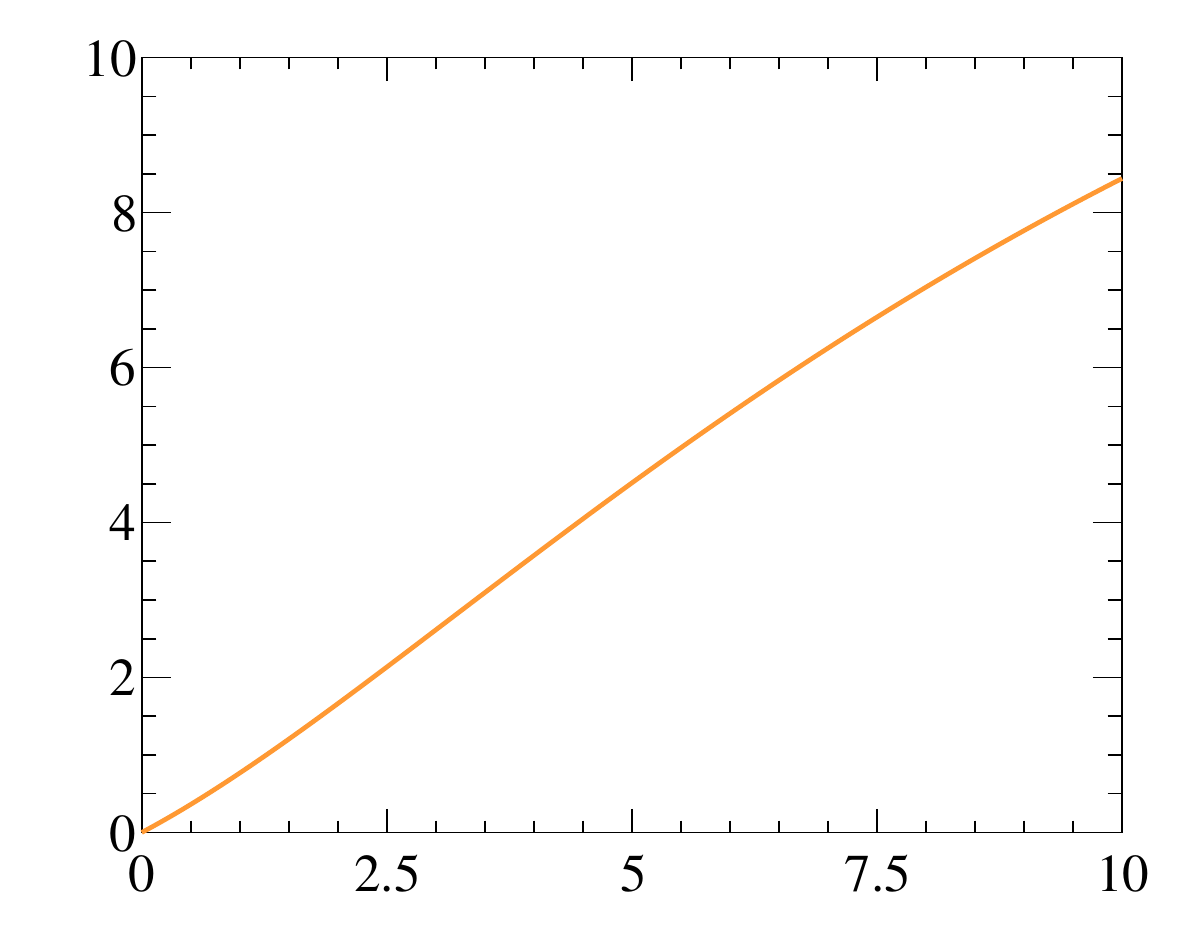}}
    \put(75,0){\includegraphics*[width=75mm,height=60mm]{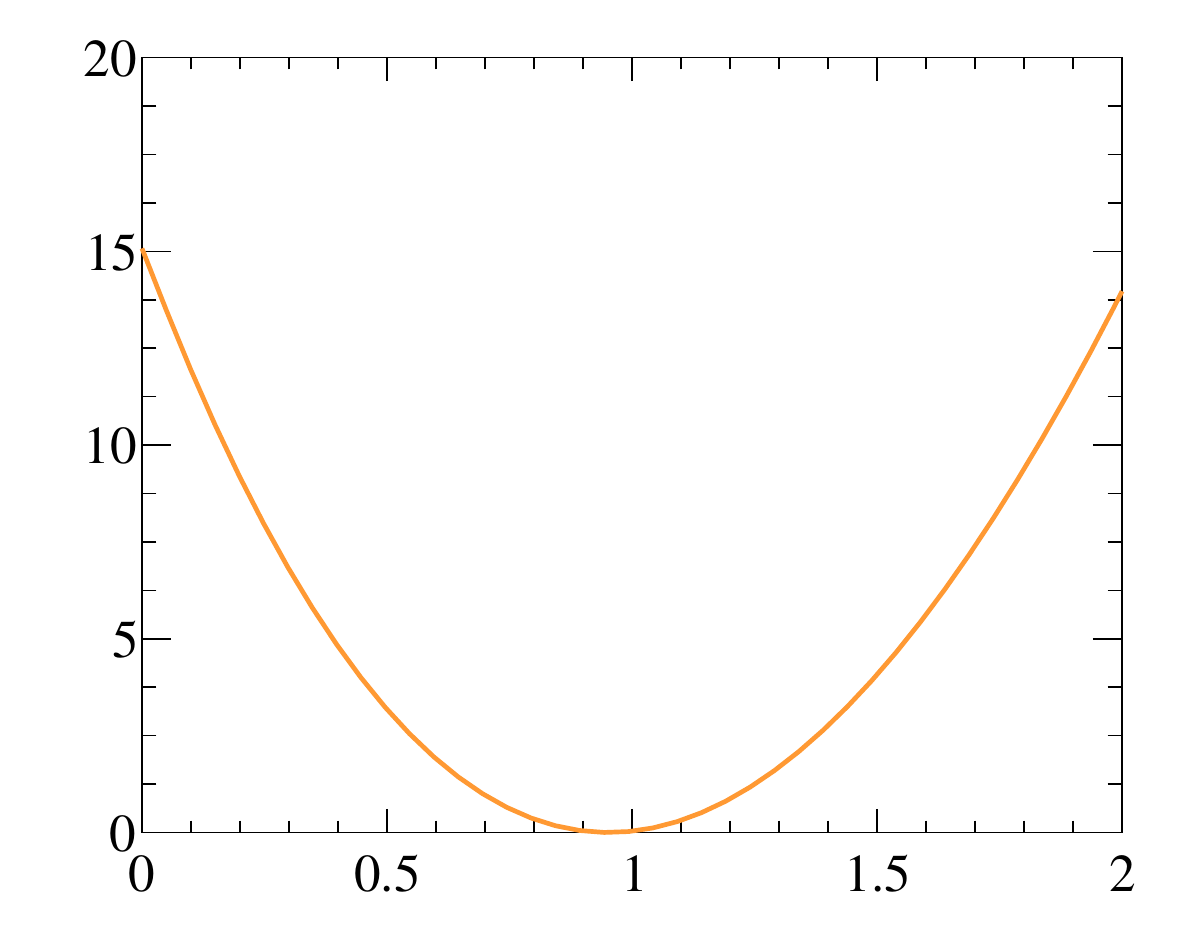}}
    
    \put(   0,41){\begin{sideways}\small{$-\Delta\log\mathcal{L}$}\end{sideways}}
    \put(  75,41){\begin{sideways}\small{$-\Delta\log\mathcal{L}$}\end{sideways}}
    
    \put(35 ,-1){$\Gamma_{\psitwod}$}
    \put(110,-1){$\Gamma_{\chiconex}$}
    
    \put( 61,-1){$\left[\!\mev\right]$}
    \put( 136,-1){$\left[\!\mev\right]$}
    
    \put( 58,51){\small\lhcb}
    \put( 134,51){\small\lhcb}
    
  \end{picture}
  \caption {\small Likelihood profiles for 
    the~Breit--Wigner width  
    of (left)\,\psitwod and (right)\,\chiconex states. }
  \label{fig:nll_plots}
\end{figure}
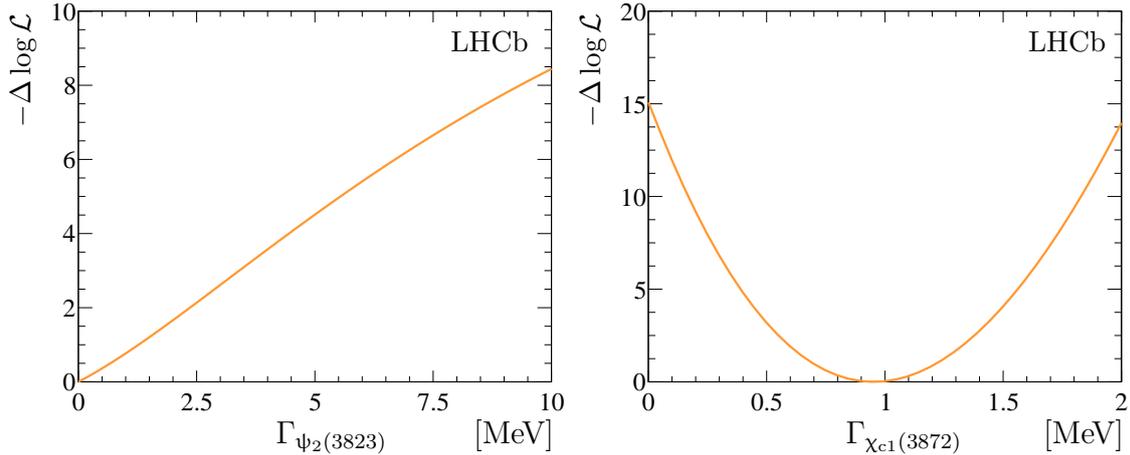

The~likelihood profiles for the~Breit--Wigner widths
of \psitwod and \chiconex states are presented in Fig.~\ref{fig:nll_plots}.
From~these profiles
the~~Breit--Wigner width of the~\chiconex~state 
is found to be inconsistent with zero by 
$5.5$~standard deviations, while for 
the~\psitwod~state the~width is consistent with zero.

%% %%%%%%%%%%%%%%%%%%%%%%%%%%%%%%%%%%%%%%%%%%%%%%%%%%%%%%%%%%%%%%%%%%%%%%%%%%%%
%% 
%% %%%%%%%%%%%%%%%%%%%%%%%%%%%%%%%%%%%%%%%%%%%%%%%%%%%%%%%%%%%%%%%%%%%%%%%%%%%%
\section{Ratios of branching fractions}

Ratios of branching fractions, $\mathcal{R}^{\mathrm{X}}_{\mathrm{Y}}$,
are defined as 
\begin{equation}
  \mathcal{R}^{\mathrm{X}}_{\mathrm{Y}} \equiv 
  \dfrac{\BR_{\decay{\Bu}{\mathrm{X}\Kp}} \times \
    \BR_{\decay{\mathrm{X}}{\jpsi\pip\pim}}}
        {\BR_{\decay{\Bu}{\mathrm{Y}\Kp}} \times \
          \BR_{\decay{\mathrm{Y}}{\jpsi\pip\pim}}}\,, \label{eq:r}    
\end{equation}
where $\mathrm{X}$, $\mathrm{Y}$ stand for either the  
\psitwod, \chiconex or \psitwos~states.
They are estimated as 
\begin{equation}
  \mathcal{R}^{\mathrm{X}}_{\mathrm{Y}} = 
  \dfrac{ N_{\decay{\Bu}{\mathrm{X}\Kp}}}
        { N_{\decay{\Bu}{\mathrm{Y}\Kp}}} \times 
        \dfrac{ \upvarepsilon_{\decay{\Bu}{\mathrm{Y}\Kp}}}
              { \upvarepsilon_{\decay{\Bu}{\mathrm{X}\Kp}}}\,, \label{eq:br_rat}
\end{equation}
where $N$ is the signal yield 
reported in Table~\ref{tab:sim_fit_res_check} 
and \eps denotes the efficiency 
of the~corresponding decay.
The~efficiency is defined as the~product of geometric acceptance, 
reconstruction, selection, hadron identification 
and trigger efficiencies, 
where each subsequent efficiency
    is defined with respect to the~previous one.
All~of the~contributions, except that of the~hadron\nobreakdash-identification efficiency, 
are determined using simulated samples.
The~hadron\nobreakdash-identification efficiency is determined
using large calibration samples
of  
\mbox{$\decay{\Dstarp}{ \left(\decay{\Dz}{\Km\pip}\right)\pip}$}, 
\mbox{$\decay{\KS}{\pip\pim}$}
and \mbox{$\decay{\Ds}{\left(\decay{\Pphi}{\Kp\Km}\right)\pip}$}
decays selected in data 
for kaons and pions~\cite{LHCb-DP-2012-003, LHCb-DP-2018-001}. 
The~ratios of the~efficiencies are determined~to~be
\begin{equation}
  \begin{aligned}
    \dfrac{\upvarepsilon_{\decay{\Bu}{\chiconex\Kp}}}
          {\upvarepsilon_{\decay{\Bu}{\psitwod\Kp}}} 
          &= 1.098 \pm 0.003\,, \\
          \dfrac{\upvarepsilon_{\decay{\Bu}{\psitwos\Kp}}}
                {\upvarepsilon_{\decay{\Bu}{\psitwod\Kp}}} 
                &= 0.778 \pm 0.003\,, \\
                \dfrac{\upvarepsilon_{\decay{\Bu}{\psitwos\Kp}}}
                      {\upvarepsilon_{\decay{\Bu}{\chiconex\Kp}}} 
                      &= 0.708 \pm 0.003\,,
  \end{aligned}
  \label{eq:eff_ratio}
\end{equation}
where the~uncertainty reflects 
the~limited size of the~simulated samples. 
Other sources  of systematic uncertainty are discussed in the~following section. 
The~ratios of the~efficiencies differ from unity
mostly due to the~different pion momentum  
spectra in the different~\mbox{$\decay{\PX_{\ccbar}}{\jpsi\pip\pim}$}~decays.

\section{Systematic uncertainty}
\label{sec:Systematics}

%% \subsection{Systematic uncertainties for the ratio of branching fractions}
Due to the~similar decay topologies, systematic uncertainties 
largely cancel in the~ratios $\mathcal{R}^{\mathrm{X}}_{\mathrm{Y}}$. 
The~remaining contributions are 
listed in Table~\ref{tab:systematics} and are discussed below.

The~systematic uncertainty related to the~signal and background
shapes is investigated using alternative
parameterisations.
%%
%% To estimate the systematic uncertainty related to the fit model, the data is fitted with
%% alternative signal and background models and the ratio $\mathcal{R}$ is computed.
%%
A~generalized Student's $t$\nobreakdash-distribution~\cite{Jackman}, 
an~Apollonios function~\cite{Santos:2013gra} and 
a~modified Novosibirsk function~\cite{PhysRevD.84.112007} 
are used as alternative models for the~\Bu~signal template.
For the~$\PX_{\cquark\cquarkbar}$~signal template, 
alternative parameterisations 
of the~mass resolution, namely 
a~symmetric variant of an~Apollonios function\cite{Santos:2013gra}, 
a~Student's $t$\nobreakdash-distribution 
and a~sum of two Gaussian functions sharing 
the~same mean are considered.
In addition, P\nobreakdash-wave and D\nobreakdash-wave 
relativistic  Breit--Wigner functions are used 
as alternative \psitwod~signal templates, and 
the~Blatt--Weisskopf barrier 
factors~\cite{Blatt:1952ije} are varied between 
$1.5$ and $5\gev^{-1}$.
The~width of the~\psitwos~state, 
fixed in the~fit, 
is varied between 270 and 302\kev~\cite{PDG2019}.
The~maximal deviations in the~ratios 
$\mathcal{R}^{\mathrm{X}}_{\mathrm{Y}}$ 
with respect to the~baseline fit model
are taken as systematic uncertainties
for each of the~systematic signal model sources.
%% 
%% The~maximal relative bias found for the ratios $\mathcal{R}$ for the signal models are determined to be
%%  \begin{itemize}
%%  	\item$\mathcal{R}^\psitwod_\chiconex:$~0.6\,(0.3)\%;
%%  	\item$\mathcal{R}^\psitwod_\psitwos:$~0.5\,(0.2)\%;
%%  	\item$\mathcal{R}^\chiconex_\psitwos:$~0.1\,(0.2)\%.
%% \end{itemize}
%% \noindent for the \jpsi\pip\pim\Kp\,(\jpsi\pip\pim) projection, 
%% which are assigned as a relative systematic uncertainty 
%% related on the signal fit model.
For~the~systematic uncertainty related to 
the~modelling of the~smooth polynomial
functions,  
%% the~non\nobreakdash-signal 
%% fit components,
pseudoexperiments
with about
    $10^7$~simulated
    events\,(approximately 100 times large than data sample
are generated
with the~baseline fit model and fitted with
alternative background models. 
In~this study the~degree of the~polynomial functions 
is varied from the~first to the~second order, 
separately for each fit component and each channel. 
In~each case  the~ratio $\mathcal{R}^{\mathrm{X}}_{\mathrm{Y}}$ 
is computed and 
the~maximal difference with respect to the~baseline 
fit model 
is taken as a~corresponding systematic uncertainty.
%% related 
%% to the~modelling 
%% of polynomial components.
%%
%%

%% found for the ratios $\mathcal{R}$ for the background model are determined to be
%% \begin{itemize}
%% 	\item$\mathcal{R}^\psitwod_\chiconex:$~2.5\%;
%% 	\item$\mathcal{R}^\psitwod_\psitwos:$~2.7\%;
%% 	\item$\mathcal{R}^\chiconex_\psitwos:$~0.2\%.
%% \end{itemize}
%% \noindent which is assigned as a relative systematic uncertainty related on the background  model.

\begin{table}[b]
	\centering
	\caption{Relative systematic uncertainties (in \%) for 
	the~ratios of branching fractions $\mathcal{R}^{\mathrm{X}}_{\mathrm{Y}}$.} 
	\vspace*{2mm}
     \begin{tabular*}{0.95\textwidth}{@{\hspace{3mm}}l@{\extracolsep{\fill}}ccc@{\hspace{2mm}}}
    Source 
		&  $\mathcal{R}^{\psitwod}_{\chiconex}$ 
		& $\mathcal{R}^{\psitwod}_{\psitwos}$ 
		& $\mathcal{R}^{\chiconex}_{\psitwos}$
    \\[1.5mm]
    \hline 
    \\[-3mm]
  Signal and background shapes & & &  \\ 
  ~~\Bu~signal template                       & $\phantom{<}0.6$     & 0.5 & 0.1 \\
  ~~$\PX_{\cquark\cquarkbar}$~signal template & $\phantom{<}0.3$     & 0.2 & 0.2 \\
  %%~~Non-signal components              & $\phantom{<}2.5$     & 2.7 & 0.2 \\
  ~~Polynomial components              & $\phantom{<}2.5$     & 2.7 & 0.2 \\
   \psitwod decay model                 & $\phantom{<}0.2$     & 0.2  & --- \\ 
  %%Efficiency corrections               & $<0.05$ & 0.16 & 0.16 \\
  Efficiency corrections               & $<0.1$ & 0.2 & 0.2 \\
  Trigger efficiency                   & $\phantom{<}1.1$     & 1.1  & 1.1 \\ 
  Data-simulation agreement            & $\phantom{<}1.0$     & 1.0  & 1.0 \\ 
  Simulation sample size               & $\phantom{<}0.3$     & 0.4  & 0.4 
  \\[1.5mm]
   \hline 
  \\[-3mm]
   Sum in quadrature                   & $\phantom{<}3.0$ & 3.2 & 1.6 
	\end{tabular*}
	\label{tab:systematics}
\end{table}

Since the~decay model for \mbox{$\decay{\psitwod}{\jpsi\pip\pim}$} 
is unknown, 
a~phase\nobreakdash-space model is used in simulation.
To~probe the associated systematic uncertainty
the~model discussed in Ref.~\cite{Wang:2015xsa} is used.
This~model accounts for the~quantum\nobreakdash-chromodynamics multipole expansion~\cite{Kuang:1988bz},
as well as the~effective description of the~coupled\nobreakdash-channel 
effects via \mbox{hadronic}\nobreakdash-loop mechanism~\cite{Liu:2006dq} 
with the~interference phase $\mathit{\Phi}$ as a~free parameter.
The~$\pip\pim$~mass spectrum and the~angular 
distributions in the~decay
strongly depend on the~phase $\mathit{\Phi}$, 
however, the~efficiency for 
the~\mbox{$\decay{\Bu}{\left(\decay{\psitwod}{\jpsi\pip\pim}\right)\Kp}$}~decays
is found to be stable.  
It~varies within 0.2\%  with respect to 
the~efficiency computed 
for the~phase\nobreakdash-space model
when the~unknown  phase $\mathit{\Phi}$ varies 
in the~range~\mbox{$-\pi\le\mathit{\Phi}<\pi$}.

An~additional uncertainty arises from 
differences between the data and  simulation,
in particular differences in the reconstruction efficiency 
of charged\nobreakdash-particle tracks.
The~track\nobreakdash-finding efficiencies obtained from 
the~simulation samples 
are corrected \mbox{using}
%%{\color{red}{\sout{data calibration samples}}}
data\nobreakdash-driven techniques~\cite{LHCb-DP-2013-002}.
%% to account for the~differences in the~data and the~simulation.
The uncertainties related to~the correction factors, 
\mbox{together}
with the~uncertainty in the~hadron\nobreakdash-identification efficiency 
due to the~finite size of 
the~calibration samples~\cite{LHCb-DP-2012-003, LHCb-DP-2018-001},
are propagated to the~ratio of total efficiencies using pseudoexperiments.

The~systematic uncertainty related to the~trigger efficiency is estimated 
using large samples of  
the~$\decay{\Bp}{\jpsi\Kp}$ 
and 
$\decay{\Bp}{\psitwos\Kp}$ decays by comparing 
the~ratios of trigger
efficiencies in data and simulation~\cite{LHCb-PAPER-2012-010}. 
The~imperfect data description by the~simulation
due to remaining effects is studied by varying the~BDT 
selection criteria in ranges that lead to
%% as much as 
$\pm 20\%$ change in the~measured efficiency. 
%% The~stability is tested by comparing the efficiency ratios within these variations. 
The~resulting variations in 
the~efficiency ratios do not exceed~$1\%$, 
which is taken as a~corresponding systematic uncertainty.
The~last systematic uncertainty 
considered for the~ratio $\mathcal{R}^{\mathrm{X}}_{\mathrm{Y}}$ is 
due to the~finite size of the~simulated samples.
%% 
%% \begin{itemize}
%%  	\item$\mathcal{R}^\psitwod_\chiconex:$~0.3\%;
%%  	\item$\mathcal{R}^\psitwod_\psitwos:$~0.4\%;
%%  	\item$\mathcal{R}^\chiconex_\psitwos:$~0.4\%.
%% \end{itemize}

For~each choice of the~fit model,  
the~statistical significance of  the~observed 
\mbox{$\decay{\Bu}{\left(\decay{\psitwod}{\jpsi\pip\pim}\right)\Kp}$}
signal is calculated from fit to data using
Wilks' theorem.
The smallest significance found is
5.1~standard deviations,
numerically close to the~value obtained
from the~baseline fit model.

%% \subsection{Systematic uncertainties for the mass measurements}
The systematic uncertainties on 
the~mass differences between the~\psitwod, 
\chiconex and \psitwos states are summarized in Table~\ref{tab:systematics_mass}.
An~important source of systematic uncertainty 
is due to the~signal and background shapes.
Different parameterisations of the~signal templates 
and non\nobreakdash-signal components, described above, 
are used as the~alternative fit models. 
The~maximal deviation in the~mass differences with respect to the~baseline results 
is assigned as the~corresponding systematic uncertainty. 
The~uncertainty in the~momentum\nobreakdash-scale calibration, 
important for mass measurements, \eg Refs.~\mbox{\cite{
LHCb-PAPER-2012-028,
LHCb-PAPER-2012-048,
LHCb-PAPER-2013-010,
LHCb-PAPER-2013-011,
LHCb-PAPER-2014-002,
LHCb-PAPER-2014-021, 
LHCb-PAPER-2014-048,
LHCb-PAPER-2015-060,
LHCb-PAPER-2016-008, 
LHCb-PAPER-2017-011, 
LHCb-PAPER-2019-005, 
LHCb-PAPER-2019-007,
LHCb-PAPER-2019-025,
LHCb-PAPER-2019-037, 
LHCb-PAPER-2019-042,
LHCb-PAPER-2019-045,
LHCb-PAPER-2020-003}}, 
largely cancels for the~mass differences.
The~associated systematic uncertainty is 
evaluated by varying the~momentum scale within its 
known uncertainty~\cite{LHCb-PAPER-2013-011} and repeating the~fit. 
The~$\jpsi\pip\pim$ mass is computed 
constraining the~mass of the~\Bu~candidate 
to the~known value, \mbox{$m_{\Bu}=5279.25 \pm 0.26 \mevcc$}~\cite{PDG2019}.
The~uncertainty on the~\Bu~meson mass %%used in the~constraint
is propagated to the~measured mass differences.  

\begin{table}[t]
	\centering
	\caption{Systematic uncertainties (in~\mevcc) for the mass splitting between 
	the~\psitwod, \chiconex and \psitwos states.}	
	\label{tab:systematics_mass} 
	\vspace*{2mm}
	 \begin{tabular*}{1\textwidth}{@{\hspace{1mm}}l@{\extracolsep{\fill}}ccc@{\hspace{1mm}}}
    Source 
		& $m_{\psitwod} - m_{\psitwos}$ 
		& $m_{\chiconex} - m_{\psitwos}$
		& $m_{\chiconex} - m_{\psitwod}$
    \\[1.5mm]
    \hline 
    \\[-3mm]
  \multicolumn{2}{l}{Signal and background shapes}   & \\
   ~~\Bu~signal template                            &  0.023  &  $\phantom{<}0.002$ &0.023 \\ 
   ~~$\PX_{\cquark\cquarkbar}$~signal template      &  0.115  &  $\phantom{<}0.005$ &0.110\\
   %% ~~Non-signal components                          &  0.070  &  
   ~~Polynomial components                          &  0.070  &  
   %% $<0.010$ 
   $\phantom{<}0.001$
   &0.070\\
   Momentum scale                                   &  0.004  &  $\phantom{<}0.009$ &0.005\\ 
   \Bu~mass uncertainty                             &  0.021  &  $\phantom{<}0.029$ &0.008    
      \\[1.5mm]
    \hline 
    \\[-3mm]
    Sum in quadrature                               &  0.138 &  $\phantom{<}0.031$  &0.133
	\end{tabular*}
	
\end{table}

%%% \subsection{Systematic uncertainties for the natural widths}
The~main source of systematic uncertainty for 
the~Breit--Wigner widths 
$\Gamma_{\psitwod}$ and 
$\Gamma_{\chiconex}$ is due to 
the~signal and background shapes.
The~maximal  $\Gamma_{\chiconex}$  deviation
%% with respect to the~baseline model 
of $0.21\mev$ is taken as the~systematic uncertainty. 
For~all the~fits, 
the~$\Gamma_{\psitwod}$~parameter is found to be consistent with zero,
and an~upper limit is obtained from analysis of
the~likelihood profile curve.
The~~maximal value of the~upper limits is conservatively
taken as the~estimate that accounts for the~systematic uncertainty 
\begin{equation}
   %%\Gamma_{\psitwod} < 5.21\,(6.56)\mev\, \text{at 90\,(95)\% C.\,L}.
\Gamma_{\psitwod} < 5.2\,(6.6)\mev\, \text{at 90\,(95)\% CL.}
\end{equation}
The~systematic uncertainty due to the~mismodelling of 
the~experimental resolution in simulation is 
accounted for 
with the~resolution scale factors $f_{\Bu}$ and $f_{\PX_{\cquark\cquarkbar}}$
and therefore is included as a~part of the~statistical uncertainty.   
%%
%%{\color{blue}{
    A~small dependency of the~scale
    factor $f_{\PX_{\cquark\cquarkbar}}$
    on the~dipion momentum
    for the~$\decay{\psitwos}{\jpsi\pip\pim}$~decay
    is reported in Ref.~\cite{LHCb-PAPER-2020-008}.
    Such effect causes a~bias
    in the~effective scale factor
    for different decays 
    due to slightly different dipion spectra.
    Such bias is  
    %% {\em{(checked and?)}}
    %% is checked
    %% convolving the~oserved scale-factor
    %% dependency~\cite{LHCb-PAPER-2020-008}
    %% with dipion spectra  
    %% and  
    found to be
    negligible with respect
    to the~statistical uncertainty
    for the~factor~$f_{\PX_{\cquark\cquarkbar}}$.
%%}}
%% 

The~analysis is carried out by neglecting any interference effects
between the~$\PX_{\ccbar}$~resonances and other components.
Such an~assumption can bias the~measurement of the~mass
and width\nobreakdash-parameters associated to the~$\PX_{\ccbar}$~states.
To~account for such interference effects 
a~full amplitude analysis is required, which is beyond 
the~scope of this study.
However,  to~estimate the~possible 
effect of this assumption 
on the~\chiconex mass and width\nobreakdash-parameters,
the~background\nobreakdash-subtracted $\jpsi\pip\pim$~mass distribution 
in the~\chiconex region is studied with 
the~\sPlot~technique used for background subtraction~\cite{Pivk:2004ty}
using the~$\jpsi\pip\pim\Kp$~mass
as the~discriminative  variable.
The~distribution is fit with a~model that accounts for
the~signal, coherent and incoherent backgrounds
\begin{equation}
    \mathcal{F}(m) = 
    {\mathcal{N}} \left( \left|  \mathcal{A}_{\mathrm{BW}}(m) + 
    b_{\mathrm{c}}(m) \,  \mathrm{e}^{i\Pdelta(m)}    \right|^2 
    \circledast \mathfrak{R} \right) 
    + b^2_{\mathrm{i}}(m)\,, \label{eq:interference}
\end{equation}
where $\mathcal{A}_{\mathrm{BW}}(m)$ is a~Breit--Wigner amplitude, 
convolved with the~mass resolution function $\mathfrak{R}$,
and ${\mathcal{N}}$~stands for a~normalisation constant. 
The~coherent and incoherent background components
$b_{\mathrm{c}}(m)$ and 
$b^2_{\mathrm{i}}(m)$ 
are parameterised with polynomial functions.
The~relative interference 
phase $\Pdelta(m)$ is taken to be constant 
for the~narrow \mbox{$3.85\le m_{\jpsi\pip\pim}<3.90\gevcc$} region,
$\Pdelta(m)\equiv\Pdelta_0$. 
An~equally good description of data is achieved for 
totally incoherent\,\mbox{($b_{\mathrm{c}}(m)\equiv0$)} 
and coherent\,\mbox{($b^2_{\mathrm{i}}(m)\equiv0$)} background hypotheses, 
as well as for any intermediate scenarios with 
the~phase $\Pdelta_0$ close to $\tfrac{\pi}{2}$.  
The~latter reflects a~high symmetry of the~observed \chiconex~lineshape.
For~all scenarios,  
variations of the~mass and width parameters
are limited to 50\kevcc and 150\kev, respectively.  

\section{Results and summary}
\label{sec:Results}

The~decay of \mbox{$\decay{\Bu}{\left(\decay{\psitwod}{\jpsi\pip\pim}\right)\Kp}$}
is observed for the~first time with 
a~significance of 5.1~standard deviations.
The~signal yield of \mbox{$137\pm26$}~candidates, 
together with \mbox{$4230\pm70$}
\mbox{$\decay{\Bu}{\left(\decay{\chiconex}{\jpsi\pip\pim}\right)\Kp}$}
and 
\mbox{$\left(81.14\pm0.29\right)\times10^3$}
\mbox{$\decay{\Bu}{\left(\decay{\psitwos}{\jpsi\pip\pim}\right)\Kp}$}~signal candidates,
allows for a~precise determination of the~ratios of the~branching fractions 
\begin{equation*}
\begin{array}{lclcl}
\mathcal{R}^{\psitwod}_{\chiconex} = 
& \dfrac{\BR_{\decay{\Bu}{\psitwod\Kp}}\times 
       \BR_{\decay{\psitwod}{\jpsi\pip\pim}}}
      {\BR_{\decay{\Bu}{\chiconex\Kp}} \times 
       \BR_{\decay{\chiconex}{\jpsi\pip\pim}}} 
       & = &   \left(3.56 \pm 0.67 \pm 0.11 \right) \times 10^{-2}\,, \\
\mathcal{R}^{\psitwod}_{\psitwos}  = 
& \dfrac{\BR_{\decay{\Bu}{\psitwod\Kp}} \times 
       \BR_{\decay{\psitwod}{\jpsi\pip\pim}}}
      {\BR_{\decay{\Bu}{\psitwos\Kp}} \times 
       \BR_{\decay{\psitwos}{\jpsi\pip\pim}}} 
       & = &   \left(1.31 \pm 0.25 \pm 0.04 \right) \times 10^{-3}\,, \\
\mathcal{R}^{\chiconex}_{\psitwos} = 
& \dfrac{\BR_{\decay{\Bu}{\chiconex\Kp}} \times 
       \BR_{\decay{\chiconex}{\jpsi\pip\pim}}}
      {\BR_{\decay{\Bu}{\psitwos\Kp}} \times 
       \BR_{\decay{\psitwos}{\jpsi\pip\pim}}} 
       & = &   \left(3.69 \pm 0.07 \pm 0.06 \right) \times 10^{-2}\,, 
\end{array}
\end{equation*}
where the~first uncertainty  is statistical 
and the~second is systematic.       
The~last ratio is in good agreement with, but significantly more precise 
than the~value 
of \mbox{$\left(4.0 \pm 0.4\right)\times10^{-2}$}, derived from Ref.~\cite{PDG2019}.
Only two ratios $\mathcal{R}^{\mathrm{X}}_{\mathrm{Y}}$
are statistically  independent.
The~non\nobreakdash-zero correlation coefficients are 
$+97\%$ for $\mathcal{R}^{\psitwod}_{\chiconex}$ and 
$\mathcal{R}^{\psitwod}_{\psitwos}$,
and $-7\%$ for 
$\mathcal{R}^{\psitwod}_{\chiconex}$ and 
$\mathcal{R}^{\chiconex}_{\psitwos}$.
The~product of branching fractions for 
the~decay via the~intermediate  \psitwod~state
is calculated to be 
\begin{equation*}
%% \BR \left(\decay{\Bu}{\psitwod\Kp}\right) \times 
%%     \BR\left(\decay{\psitwod}{\jpsi\pip\pim}\right) 
\BR_{\decay{\Bu}{\psitwod\Kp}} \times 
\BR_{\decay{\psitwod}{\jpsi\pip\pim}} 
%% =  \left(2.82125 \pm 0.538407  
%% \pm 0.0861 \pm 0.10288\right)\times 10^{-7}\,,
=  \left(2.82  \pm 0.54  
\pm 0.09 \pm 0.10\right)\times 10^{-7}\,,      
\end{equation*}
where the last uncertainty is due to 
the knowledge of the~branching fractions 
for \mbox{$\decay{\Bu}{\psitwos\Kp}$} and 
\mbox{$\decay{\psitwos}{\jpsi\pip\pim}$} decays~\cite{PDG2019}.
Combined with 
the~calculated 
value of 
%% \mbox{$\BR\left(\decay{\psitwod}{\jpsi\pip\pim}\right)$}~\cite{Xu:2016kbn}
%% $\BR_{\decay{\psitwod}{\jpsi\pip\pim}}$~\cite{Xu:2016kbn},
$\BR_{\decay{\psitwod}{\jpsi\Ppi\Ppi}}$~\cite{Xu:2016kbn}
this yields 
$\BR_{\decay{\Bu}{\psitwod\Kp}}=(1.24\pm0.25)\times10^{-6}$. 
This is smaller but more precise
than the~value of 
$(2.1\pm0.7)\times10^{-5}$ derived from
the~measurement of 
$\BR_{\decay{\Bu}{\psitwod\Kp}} \times 
\BR_{\decay{\psitwod}{\chicone\g}}
=(9.7\pm2.8\pm1.1)\times10^{-6}$
by the~Belle collaboration~\cite{Bhardwaj:2013rmw}
%% $\BR_{\decay{\Bu}{\psitwod\Kp}} \times 
%% \BR_{\decay{\psitwod}{\chicone\g}}
%% =(9.7\pm2.8\pm1,1)\times10^{-6}$~\cite{Bhardwaj:2013rmw},
and the~estimate for 
$\BR_{\decay{\psitwod}{\chicone\g}}$~\cite{Xu:2016kbn}.
Within a~factorization approach the~branching fraction
for the~decay  \mbox{$\decay{\Bu}{\psitwod\Kp}$}
vanishes, and a~large value for 
this~branching fraction requires  a~large 
contribution of the~$\D{}^{(*)+}_{\squark}\Dbar{}^{(\ast)0}$~rescattering 
amplitudes in the~\decay{\Bu}{\ccbar\Kp}~decays~\cite{Xu:2016kbn}. 
This~measurement of 
the~branching fraction for 
the~$\decay{\Bu}{\psitwod\Kp}$~decay
allows for a~more precise estimation of   the~role 
of the~$\D_{\squark}^{(*)+} \Dbar{}^{(\ast)0}$~rescattering 
mechanism~\cite{Xu:2016kbn}.

Using a~Breit$-$Wigner parameterisation, 
the~mass differences between the~\psitwod, \chiconex 
and \psitwos states are found to be 
\begin{eqnarray*}
%% \textcolor{red}{m_\chiconex - m_\psitwos} &= &  185.48831680539496 \pm  0.06230793653235075 \\
m_\chiconex - m_\psitwod &= &  \phantom{0}47.50  \pm 0.53 \pm 0.13 \mevcc\,, \\
m_\psitwod  - m_\psitwos &= &  137.98 \pm 0.53 \pm 0.14 \mevcc\,, \\
m_\chiconex - m_\psitwos &= &  185.49 \pm 0.06 \pm 0.03 \mevcc\,. 
%% 185.4883168 \pm 0.0623070 \pm 0.30 
\end{eqnarray*}
Only two from 
three mass differences are independent.
Two~non\nobreakdash-zero correlation coefficients 
are $-93\%$
for 
$m_{\chiconex}-m_{\psitwod}$ and 
$m_{\psitwod}-m_{\psitwos}$ and 
$+10\%$ 
for 
$m_{\chiconex}-m_{\psitwod}$ and 
$m_{\chiconex}-m_{\psitwos}$.

The~Breit--Wigner width of the~\chiconex~state is found to be
\begin{equation*}
\Gamma_{\chiconex} =0.96 ^{\,+\,0.19}_{\,-\,0.18} \pm0.21 \mev\,,
\end{equation*}
which is inconsistent with zero by 5.5~standard deviations.
The~width of the~\psitwod~state is found to be consistent with zero
and an~upper limit at 90\%\,(95\%) confidence level is set~at
\begin{equation*}
   \Gamma_{\psitwod} < 5.2\,(6.6)\mev\,. 
 \end{equation*}
The~value of the~Breit--Wigner width $\Gamma_{\chiconex}$
agrees well with the~value from the~analysis of 
a~large sample of \mbox{$\decay{\chiconex}{\jpsi\pip\pim}$}
decays from the~inclusive decays
of beauty hadrons~\cite{LHCb-PAPER-2020-008}.
Using the~known value of the~\psitwos mass~\cite{PDG2019}, 
the~Breit$-$Wigner masses 
for the~\psitwod and \chiconex~states 
are computed to be 
\begin{eqnarray*}
m_{\psitwod}   & = &  3824.08 \pm 0.53 \pm 0.14 \pm 0.01 \mevcc \,, \\
m_{\chiconex}  & = &  3871.59 \pm 0.06 \pm 0.03 \pm 0.01 \mevcc\,, 
%%  3871.5853 \pm 0.0623079 \pm 0.030 \om 0.010
\end{eqnarray*}
where the last uncertainty is due to the~knowledge of the~\psitwos~mass. 
These are the~most precise measurements of these masses.

The~mass difference between \chiconex and \psitwos states
is more precise than the~average reported in Ref.~\cite{PDG2019}.
It~also agrees well with the measurement from Ref.~\cite{LHCb-PAPER-2020-008}.
Taking into account a~partial overlap of the~data sets and 
correlated part of systematic 
uncertainty, the~LHCb average  mass difference and 
the~mass of the~\chiconex~state are
\begin{eqnarray*} 
%% \left. m_{\chiconex} - m_{\psitwos} \right|_{\mathrm{LHCb}}& = & \phantom{0}185.542 \pm 0.060 \mevcc\,, \\ 
%% \left. m_{\chiconex}\right|_{\mathrm{LHCb}}  & = & 3871.639 \pm 0.060 \pm 0.010 \mevcc\,, 
\left. m_{\chiconex} - m_{\psitwos} \right|_{\mathrm{LHCb}}& = & \phantom{0}185.54 \pm 0.06 \mevcc\,, \\ 
\left. m_{\chiconex}\right|_{\mathrm{LHCb}}  & = & 3871.64 \pm 0.06 \pm 0.01 \mevcc\,, 
\end{eqnarray*}
where the~second uncertainty is due to the knowledge of the~\psitwos~mass. 
The difference between the $m_{\chiconex}$ mass, determined from the~Breit--Wigner fit, 
and the~$\Dz\Dstarz$~threshold
$\delta E\equiv \left(\m_{\Dz}+m_{\Dstarz}\right)c^2- m_{\chiconex}c^2$ 
is computed to be 
\begin{eqnarray*}
%% \delta E ~~~~~~                         & = &  119 \pm 128    \kev\,, \\
%% \left.  \delta E \right|_{\mathrm{LHCb}} & = &\phantom{0}66 \pm 124 \kev     \,,
\delta E ~~~~~~                          & = &  0.12 \pm 0.13    \mev\,, \\
\left.  \delta E \right|_{\mathrm{LHCb}} & = &  0.07 \pm 0.12 \mev     \,,
\end{eqnarray*}
where the~first value corresponds to the measurement performed in this analysis, 
while the~second one is an average with results from Ref.~\cite{LHCb-PAPER-2020-008}.
A ~value of \mbox{$3871.70\pm0.11\mevcc$} is taken for 
the~threshold \mbox{$\m_{\Dz}+m_{\Dstarz}$}, calculated from Ref.~\cite{PDG2019,LHCb-PAPER-2020-008},
accounting for the~correlation due to the~knowledge of 
the~charged and neutral kaon masses between the~measurements. 
The~uncertainty on $\delta E$ is now dominated 
by the~knowledge of kaon masses.
These are the~most precise measurements of 
the~$\chiconex$~mass and $\delta E$~parameter.

\section*{Acknowledgements}
%
% These Acknowledgements valid from 3-May-2019
%
\noindent 
We~thank X.~Liu for the~useful discussion on 
the~\mbox{$\decay{\psitwod}{\jpsi\pip\pim}$}
and \mbox{$\decay{\Bu}{\psitwod\Kp}$}~decays 
and A.V.~Luchinsky for providing
us with the~code for modelling
the~$\decay{\psitwod}{\jpsi\pip\pim}$~decays.
We~express our gratitude to our colleagues in the~CERN
accelerator departments for the~excellent performance of the LHC. 
We~thank the technical and administrative staff at the LHCb
institutes.
We~acknowledge support from CERN and from the national agencies:
CAPES, CNPq, FAPERJ and FINEP\,(Brazil); 
MOST and NSFC\,(China); 
CNRS/IN2P3\,(France); 
BMBF, DFG and MPG\,(Germany); 
INFN\,(Italy); 
NWO\,(Netherlands); 
MNiSW and NCN\,(Poland); 
MEN/IFA\,(Romania); 
MSHE\,(Russia); 
MinECo\,(Spain); 
SNSF and SER\,(Switzerland); 
NASU\,(Ukraine); 
STFC\,(United Kingdom); 
DOE NP and NSF\,(USA).
%5
We~acknowledge the~computing resources that are provided by CERN, 
IN2P3\,(France), 
KIT and DESY\,(Germany), 
INFN\,(Italy), 
SURF\,(Netherlands),
PIC\,(Spain), 
GridPP\,(United Kingdom), 
RRCKI and Yandex LLC\,(Russia), 
CSCS\,(Switzerland), 
IFIN\nobreakdash-HH\,(Romania), 
CBPF\,(Brazil),
PL\nobreakdash-GRID\,(Poland) and
OSC\,(USA).
We~are indebted to the~communities behind the~multiple open-source
software packages on which we depend.
Individual groups or members have received support from
AvH Foundation (Germany);
EPLANET, Marie Sk\l{}odowska\nobreakdash-Curie Actions and ERC\,(European Union);
ANR, Labex P2IO and OCEVU, and 
R\'{e}gion Auvergne\nobreakdash-Rh\^{o}ne\nobreakdash-Alpes\,(France);
Key Research Program of Frontier Sciences of CAS, CAS PIFI, and 
the Thousand Talents Program\,(China);
RFBR, RSF and Yandex~LLC\.(Russia);
GVA, XuntaGal and GENCAT\,(Spain);
the Royal Society
and the~Leverhulme Trust\,(United Kingdom).

%% \clearpage
%% \input{supplementary-app}

%%\usetikzlibrary{patterns}

\clearpage
\addcontentsline{toc}{section}{References}
\bibliographystyle{LHCb}
\bibliography{main,standard,LHCb-PAPER,LHCb-CONF,LHCb-DP,LHCb-TDR}
 
\newpage

% LHCb collaboration author list
% Data extracted on May 22nd, 2020 at 2:56pm for reference date 07-Apr-2020
\centerline
{\large\bf LHCb collaboration}
\begin
{flushleft}
\small
R.~Aaij$^{31}$,
C.~Abell{\'a}n~Beteta$^{49}$,
T.~Ackernley$^{59}$,
B.~Adeva$^{45}$,
M.~Adinolfi$^{53}$,
H.~Afsharnia$^{9}$,
C.A.~Aidala$^{82}$,
S.~Aiola$^{25}$,
Z.~Ajaltouni$^{9}$,
S.~Akar$^{64}$
% ...

{\footnotesize \it

% ...
$ ^{9}$Universit{\'e} Clermont Auvergne, CNRS/IN2P3, LPC, Clermont-Ferrand, France\\
% ...
$ ^{25}$INFN Sezione di Milano, Milano, Italy\\
% ...
$ ^{31}$Nikhef National Institute for Subatomic Physics, Amsterdam, Netherlands\\
% ...
$ ^{45}$Instituto Galego de F{\'\i}sica de Altas Enerx{\'\i}as (IGFAE), Universidade de Santiago de Compostela, Santiago de Compostela, Spain\\
% ...
$ ^{49}$Physik-Institut, Universit{\"a}t Z{\"u}rich, Z{\"u}rich, Switzerland\\
% ...
$ ^{53}$H.H. Wills Physics Laboratory, University of Bristol, Bristol, United Kingdom\\
% ...
$ ^{59}$Oliver Lodge Laboratory, University of Liverpool, Liverpool, United Kingdom\\
% ...
$ ^{64}$University of Cincinnati, Cincinnati, OH, United States\\
% ...
$ ^{67}$Syracuse University, Syracuse, NY, United States\\
% ...
$ ^{82}$University of Michigan, Ann Arbor, United States, associated to $^{67}$\\
% ...
\medskip
}
\end{flushleft}

\end{document}